\documentclass[aps,pra,twocolumn,showpacs,amsmath,amssymb,floatfix,superscriptaddress,notitlepage,longbibliography]{revtex4-2}

\usepackage{color}
\usepackage{bbm}
\usepackage{graphicx}
\usepackage[utf8]{inputenc}

\usepackage{comment}
\usepackage{graphicx}
\usepackage{epstopdf}
\usepackage{amsthm}
\usepackage{amsmath}
\usepackage{empheq}
\usepackage{physics}
\usepackage{amssymb}
\usepackage[usenames,dvipsnames]{xcolor}
\usepackage{cases}
\usepackage{latexsym}
\usepackage[colorlinks=true,linkcolor=blue, citecolor=blue, urlcolor=blue, bookmarks]{hyperref}
\usepackage[title]{appendix}
\usepackage{enumitem}
\usepackage{scalerel, wasysym}
\usepackage{ifthen,amsxtra,amsfonts,dsfont,bm,tikz,CircuitTikz}

\newcommand{\1}{\mathbbm{1}}

\newtheorem{property}{Property}
\theoremstyle{plain}

\newcommand{\be}{\begin{equation}}
\newcommand{\ee}{\end{equation}}

\date{\today}

\begin{document}

\author{Jonathon Riddell}
\affiliation{School of Physics and Astronomy, University of Birmingham, Edgbaston, Birmingham, B15 2TT, UK}

\author{Katja Klobas}
\affiliation{School of Physics and Astronomy, University of Birmingham, Edgbaston, Birmingham, B15 2TT, UK}

\author{Bruno Bertini}
\affiliation{School of Physics and Astronomy, University of Birmingham, Edgbaston, Birmingham, B15 2TT, UK}

\title{Quantum State Designs from Minimally Random Quantum Circuits}

\begin{abstract}
Random many-body states and matrices are both a useful tool to model certain physical systems and an important asset for quantum computation. Realising them, however,  generally requires an exponential (in system size) amount of resources. Recent research has presented a way out by showing that one can generate random states and matrices, or more precisely a controlled approximation of them, by applying a quantum circuit built in terms of few-body unitary gates. Most of this research has been focussed on the case of quantum circuits composed by fully random unitary gates. Here we consider what happens for circuits that, instead, involve a minimal degree of randomness. Specifically, we concentrate on two different settings: (a) brickwork quantum circuits with a single one-qudit random matrix at a boundary; (b) brickwork quantum circuits with fixed interactions but random one-qudit gates everywhere. We show that, for any given initial state, (a) and~(b) produce a distribution of states approaching the Haar distribution in the limit of large circuit depth. More precisely, we show that the moments of the distribution produced by our circuits can approximate the ones of the Haar distribution in a depth proportional to the system size. Interestingly we find that in both Cases~(a) and~(b) the relaxation to the Haar distribution occurs in two steps --- this is in contrast with what happens in fully random circuits. Moreover, we show that choosing appropriately the fixed interactions, for example taking the local gate to be a dual-unitary gate with high enough entangling power, minimally random circuits produce a Haar random distribution more rapidly than fully random circuits. In particular, dual-unitary circuits with maximal entangling power --- i.e.\ perfect tensors --- appear to provide the optimal quantum state design preparation for any design number. Our results also suggest that the minimally random settings~(a) and~(b) give an optimal framework to prepare unitary designs, i.e., random unitary matrices.
\end{abstract}

\maketitle

\section{Introduction}

Recent years have seen increasing interest devoted to the realisation of many-body random states in extended quantum systems. This is because these objects provide useful models for the state of real physical systems in certain conditions --- most notably black holes~\cite{page1993information, hayden2007black} --- and represent useful resources for quantum computation and for the concrete realisation of quantum technologies, e.g., they can be used for  efficient implementations of quantum state tomography~\cite{huang2020predicting} and benchmarking~\cite{eisert2020quantum}. 

The basic problem that the researchers had to solve is that realising many-body random states, or, equivalently, many-body random unitary matrices, is \emph{hard}, i.e., requires an amount of resources scaling \emph{exponentially} with the number of microscopic constituents~\cite{knill1995approximation}. The proposed solutions made use of the concept of 
\emph{pseudorandomness}~\cite{emerson2003pseudo}, i.e., instead of attempting to construct  a bona fide random state one can try to reproduce a state that `looks' random when measuring only its first $k$ moments --- a state with this property is referred to as a quantum state $k$-design~\cite{ambanis2007quantum}. Analogously, a unitary matrix that `looks' random only up to its $k$-th moment is called unitary $k$-design~\cite{gross2007evenly}.

Crucially, contrary to true random objects, $k$-designs admit efficient realisations in terms of local quantum circuits~\cite{harrow2009random, brandao2013exponential, brandao2016local, brandao2016efficient, hunterjones2019unitary, haferkamp2022random, harrow2023approximate, haferkamp2023efficient, hearth2024unitary, schuster2024random, chen2024incompressibility, laracuente2024approximate, liu2024unitary, turkeshi2024hilbert, christopoulos2024universal, lami2025anticoncentration, sauliere2025universal, belkin2025absence}, i.e., they can be systematically approximated by repeatedly applying unitary operations that only couple a small number of qudits at a time. Specifically, a seminal result has been to show that one can realise unitary $k$-designs using local circuits whose depth is only linear in the number of qudits of the system~\cite{brandao2016local, brandao2016efficient}. In fact, very recent breakthroughs have shown that, one can achieve a specific form of approximate $k$-designs~\cite{brandao2016local} in a depth scaling poly-logarithmically with the number of constituent qudits~\cite{laracuente2024approximate, schuster2024random}.

Nearly all of this research, however, has been focussed on the case where the quantum circuits realising the approximate $k$-designs are fully random (or are fully random modulo certain symmetries~\cite{hearth2024unitary, liu2024unitary}). A natural question is then whether one can find $k$-designs when the amount of randomness is \emph{reduced to a minimum}. For instance, when the random operations are restricted on a single qudit, while the rest of the circuit only involves \emph{fixed} (non-random) ones. 

This question is particularly compelling for two main reasons. First, by optimising the fixed operations one can hope to generate $k$-designs \emph{more efficiently} than in fully random settings. In fact, the potential advantage is two-fold. On the one hand one would need to generate far less instances of the random circuit --- the exponential scaling of the number of samples with the system size is avoided providing a huge saving of resources. On the other, by selecting an optimal circuit, one can hope to generate $k$-designs \emph{faster} than the random circuit, i.e., requiring lower depth. This can be seen as a violation occurring  in quantum systems of the so called `censoring inequalities'~\cite{harrow2023approximate, schuster2024random, belkin2025absence}, which formalise the intuitive idea that more local randomness corresponds to more global randomness. Second, in the minimally random setting the emergence of $k$-designs can be seen as a fascinating novel form of \emph{universality} of the time evolution. Indeed, in the putative family of minimally random circuits producing $k$-designs the dynamics becomes eventually insensitive to all the microscopic details (of both interactions and initial state). This would not only be the case for local observables as in regular thermalising systems~\cite{calabrese2016introduction, eisert2015quantum} but also for their multilinear functions up to order $k$, hence displaying a form of thermalisation that is much stronger than the typical one --- a similar stronger form of thermalisation has been recently observed in the context of \emph{projected ensembles}~\cite{choi2023preparing, cotler2023emergent, ho2022exact, ippoliti2022solvable, Ippoliti2023, chang2024deep, mark2024maximum}.

Here we initiate a systematic investigation of the aforementioned question. We first focus on the setting where a one-qudit random matrix is applied at the boundary of a one-dimensional brickwork circuit using a combination of numerical and analytical approaches. We consider two different classes of random one-qudit matrices: random Paulis --- drawn with equal probability --- and random unitary matrices --- drawn from the Haar distribution. First, by adapting a theorem from Ref.~\cite{Ippoliti2023}, we show that for almost all choices of local gates of the quantum circuit this setting produces $k$-designs for sufficiently large times (i.e.\ circuit depths). Then we move on to characterise the time-scale of this process showing that, similarly to the fully random case~\cite{brandao2016efficient, hunterjones2019unitary, haferkamp2022random}, $k$-designs are approached exponentially fast in time. In fact, we find that if the local gates are close to being dual-unitary (DU)~\cite{bertini2019exact} and have sufficiently high entangling power the time scale for this process is strictly shorter than the Haar random circuit. Finally, we observe that, contrary to what happens for Haar random circuits, the generation of quantum state $k$-designs generically occurs in \emph{two steps}.

We then move on to investigate a `{more random}' setting that is still `less random' than the Haar circuit. Our point of departure is that the two-step relaxation we identified in the approach to $k$-designs of minimally random circuits has recently been found in the dynamics of purity~\cite{bensa2021fastest,znidaric2022solvable,znidaric2023phantom} and out-of-time-ordered correlators~\cite{bensa2022two,znidaric2023two} in finite circuits of both fixed~\cite{znidaric2023two} and random gates~\cite{bensa2021fastest,znidaric2022solvable,znidaric2023phantom,bensa2022two}. The random circuits showing two-step relaxation, however, only involve \emph{one-site} random operations while the two-qudit operation (the interaction) is fixed --- some aspects of $k$-design generation in such systems have recently been studied in Ref.~\cite{suzuki2024global}, which dubbed them \emph{structured} random circuits. This behaviour was explained in Ref.~\cite{Jonay2024} (see also Ref.~\cite{jonay2024twostage}) by noting that both fixed and structured random gates can produce `magnon-like' excitations in the multi-replica space that are not present in a fully random circuit. This suggests that structured random circuits should behave more similarly to fixed systems also in the generation of $k$-designs. We verify this hypothesis finding that all the behaviours observed in our minimally random circuits are also detected in structured random circuits. In fact, we find that random DU circuits with maximal entangling power show the fastest generation of $k$-designs. 

The rest of the paper is laid out as follows. In Sec.~\ref{sec:setting} we describe in detail both settings considered in this paper --- minimally random circuits and structured random circuits --- and the main quantity of interest: the Frobenius distance between the $k$-moment density matrix of the brickwork circuit and the global Haar random one. In Sec.~\ref{sec:numericalsurvey} we present a numerical survey of the evolution of this distance considering different choices of $k$ and gate parameters. In Sec.~\ref{sec:analysis} we present an analytical characterisation of these behaviours with special focus on the DU case, where more precise statements can be made. Finally, Sec.~\ref{sec:conclusions} contains a discussion of the results presented and our conclusions.

\section{Minimally random quantum circuits}
\label{sec:setting}

As discussed in the introduction, we are interested in the class of states generated by minimally random quantum circuits. In particular, we consider unitary circuits acting on $2L$ qudits (${d\geq2}$ internal states) with open boundary conditions and a brickwork structure of the form
\be
\label{eq:circuit}
\begin{aligned} 
        &\mathcal{U} = \mathcal{U}_e \mathcal{U}_o, \\  
        &\mathcal{U}_e = \prod_{n=1}^{L} U^{(2n)}_{2n,2n+1}, & &\mathcal{U}_o = \prod_{n=1}^{L-1} U^{(2n+1)}_{2n+1,2n+2},
\end{aligned}
\ee
such that the time evolution of a state is written as 
\begin{equation}
    |\psi(t)\rangle = \mathcal{U}^t |\psi\rangle. 
\end{equation}
In Eq.~\eqref{eq:circuit} the operator $U^{(n)}_{a,b}$ acts non-trivially, as the matrix $U^{(n)}\in U(d)$, only on the qudits at position $a$ and $b$ and we have labeled the local gates with an additional superscript index to stress that these need-not be the same. We consider inserting randomness in two different ways
\begin{enumerate}[label=(\alph*),ref=(\alph*)]
  \item \label{CaseA} Random one-qudit unitaries on one of the boundaries (independent in time) and \emph{fixed} gates everywhere else.
  \item \label{CaseB} Random (and independent in the space-time) one-site gates everywhere but \emph{fixed} two-site interactions.  
\end{enumerate}
In Case~\ref{CaseA} one can express Eq. \eqref{eq:circuit} at a given time step $t$ as
\begin{equation}
 \mathcal{U}(\alpha_t) = \mathcal{U}_e \mathcal{U}_o \cdot (\alpha_t)_{2L},
\end{equation}
where $\alpha_t \in S$ is a random one-site unitary matrix in the set $S$ independently drawn with probably $\mu_S(\alpha_t)$~\footnote{With obvious modifications this also applies for continuous probability measures.}. Case~\ref{CaseB}, instead, involves adding random one-site unitaries to every gate
\begin{equation}\label{eq:refBdeftwositeU}
   \!\!\! U^{(n)}_{n,n+1}\!=\! (\alpha^{(n)}_t \!\!\otimes \!\alpha^{(n+1)}_t) \bar U^{(n)}_{n,n+1},
\end{equation}
where $\alpha^{(n)}_t$ and $\alpha^{(n+1)}_t$ are independently drawn from $S$ and $\bar U^{(n)}_{n,n+1}$ is a fixed two-site unitary. Therefore, the time evolution operator at a given $t$ depends on 
\be
\vec{\alpha}_t=(\alpha^{(1)}_{2t-1}, \ldots,\alpha^{(2L)}_{2t-1}, \alpha^{(1)}_{2t}, \ldots,\alpha^{(2L)}_{2t})
\ee
random unitary matrices. 

To treat both Cases~\ref{CaseA} and~\ref{CaseB} simultaneously we write $\mathcal{U}(\vec\alpha_t)$ where $\vec\alpha_t$ can either be a one-dimensional or $4L$-dimensional vector of unitary random matrices. Case~\ref{CaseB} involves considerably more randomness compared to \ref{CaseA} but all of the entangling power of the circuit is contained in fixed, non-random two-local gates. We will see that this leads it to be closer to Case~\ref{CaseA} than to a circuit with fully random interactions, like those studied in Refs.~\cite{harrow2009random, brandao2013exponential, brandao2016local, brandao2016efficient, hunterjones2019unitary, haferkamp2022random, haferkamp2023efficient, hearth2024unitary, schuster2024random, chen2024incompressibility, laracuente2024approximate}.  

Given an initial state $|\psi\rangle$ we label its path through the random circuit with $\boldsymbol{\alpha}_t = \{\vec{\alpha}_1,\vec{\alpha}_2, \ldots \vec{\alpha}_t\}$, i.e., we write the state at time $t$ as $\ket{\psi({\boldsymbol{\alpha}_t})}$. Therefore we denote the ensemble produced by our random circuits by  
\begin{equation} 
\label{eq:ensemble}
    \mathcal{E} = \{ \mu_S({\boldsymbol{\alpha}_t}),\quad \ket{\psi({\boldsymbol{\alpha}_t})} \},
\end{equation}
where $\mu_S({\boldsymbol{\alpha}_t})$ is the probability of producing the state $\ket{\psi({\boldsymbol{\alpha}_t})}$. Our independence condition implies that $\mu_S({\boldsymbol{\alpha}_t})$ is the product of the probability measures for the individual $\alpha_t^{(n)}$. To fix the ideas here we consider three specific choices of $d$ (local dimension), $S$, and $\mu_S$ 
\begin{enumerate}[label={(a.\arabic*)},ref={(a.\arabic*)}]
  \item \label{CaseA1} $d=2$;   $S = \{I, \sigma^x, \sigma^y, \sigma^z\}$;   $\mu_S({\alpha})={1}/{4}$
  \item \label{CaseA2} $d$ {generic}; $S = U(d)$; $\mu_S({\alpha})=\mu_H({\alpha})$
\end{enumerate}
\begin{enumerate}[label={(b.\arabic*)},ref={(b.\arabic*)}]
  \item \label{CaseB1} $d$ {generic}; $S = U(d)^{4L}$; $\mu_S(\vec{\alpha})=\prod_{j=1}^{4L} \mu_H({\alpha_j})$
\end{enumerate}
where $\mu_H({\alpha})$ is the Haar measure of $U(d)$. We stress that the first two choices pertain to Case~\ref{CaseA} while the third to Case~\ref{CaseB}. 

We note that, by exchanging the roles of space and time, the ensemble produced by~\ref{CaseA1} becomes very similar to the projected ensemble considered in Refs.~\cite{claeys2022emergent, Ippoliti2023}. Indeed, the application of a random Pauli becomes a measurement in the Pauli basis. Our setting, however, can produce more general ensembles because --- contrary to Refs.~\cite{claeys2022emergent, Ippoliti2023} --- our local gates are \emph{not} forced to be DU. Moreover, Refs.~\cite{claeys2022emergent, Ippoliti2023} only studied the analogue of the infinite-time limit, while here we are interested in the  finite-time dynamics. Instead, as we mentioned in the introduction, random state generation in the setting~\ref{CaseB1} has recently been studied in Ref.~\cite{suzuki2024global}, which dubbed it \emph{structured} random unitary circuit. Ref.~\cite{suzuki2024global}, however, only considered 2-design preparation focussing on a \emph{special family} of local gates fulfilling a `solvability condition' such that the averaged evolution is written in terms of free fermions. In our paper we make no such restrictions: we consider generic gates and study the generation of $k$-designs with $k\geq1$. Moreover, one of our main goals is to characterise Case~\ref{CaseA} circuits, which are harder to analyse but arguably more interesting as they involve the minimal amount of randomness.

Our main question is whether the distribution in Eq.~\eqref{eq:ensemble} approaches the Haar distribution in the limit of large $t$ and, in case, how it does so. To this end we study the time evolution of the moments of the distribution, which are captured by the mixed state 
\begin{equation}
\label{eq:kmoment}
    \rho_t^{(k)} = \mathbb{E}\left[ \left( \ketbra{\psi({\boldsymbol{\alpha}_t})}{\psi({\boldsymbol{\alpha}_t})}\right)^{\otimes k} \right],
\end{equation}
where the average is taken over the measure $\mu$, i.e.
\be
\mathbb{E}[f({\boldsymbol{\alpha}_t})] = \sum_{\boldsymbol{\alpha}_t\in S}f({\boldsymbol{\alpha}_t})\mu_S({\boldsymbol{\alpha}_t}) \,.
\ee
To quantify the convergence we investigate the two-norm distance of $\rho_t^{(k)}$ to the Haar ensemble
\begin{equation} 
\label{eq:2norm}
    \Delta_2^{(k)}(t) = \frac{\| \rho_t^{(k)} - \rho_H^{(k)}\|_2^2}{\| \rho_H^{(k)}\|_2^2} ,
\end{equation}
where $\rho^{(k)}_H$ is written as Eq.~\eqref{eq:kmoment} with $\mu$ replaced by the Haar measure. Here we have chosen the two-norm as it facilitates the analytical treatment. For instance, it allows us to express the distance between the ensembles in terms of \emph{frame potentials}~\cite{renes2004symmetric, roberts2017chaos}
\begin{equation} \label{eq:framepot}
  F^{(k)} = \sum_{\boldsymbol{\alpha}_t,\boldsymbol{\beta}_t}
  \mu_{\boldsymbol{\alpha}_t} \mu_{\boldsymbol{\beta}_t} 
  \left|\braket{\psi_{\boldsymbol{\alpha}_t}}{\psi_{\boldsymbol{\beta}_t}}\right|^{2k} = \tr[\smash{\rho^{(k)\,2}}], 
\end{equation}
as follows 
\begin{equation}
    {\Delta_2^{(k)}} = \frac{F^{(k)}}{F_H^{(k)}} - 1,
\end{equation}
where the frame potential for the Haar ensemble is known to be~\cite{roberts2017chaos} 
\begin{equation} 
\label{eq:Haarmoments}
    F_H^{(k)} = \binom{\mathcal N+k-1}{k}^{-1},
 \end{equation}
where we denoted the dimension of the Hilbert space by $\mathcal N = d^{2L}$. Note that Eq.~\eqref{eq:Haarmoments} implies that the frame potentials $F^{(k)}$ are not self-averaging for the Haar ensemble, i.e., 
\be
\lim_{\mathcal N\to 0 }\frac{F_H^{(2k)}-\left (F_H^{(k)}\right )^2}{\left (F_H^{(k)}\right)^2} = \frac{(2k!)-(k!)^2}{(k!)^2} \neq 0\,. 
\ee
We also stress that Eq.~\eqref{eq:framepot} can be interpreted as the $k$-th moment of the distribution overlap magnitudes squared, i.e., $\{|\braket{\psi_{\boldsymbol{\alpha}_t}}{\psi_{\boldsymbol{\beta}_t}}|^2\}$. One can then show that, for large $L$, the condition $F^{(k)}=F_H^{(k)}$ for all $k$ is equivalent to say that $\{|\braket{\psi_{\boldsymbol{\alpha}_t}}{\psi_{\boldsymbol{\beta}_t}}|^2\}$ follow the well-known Porter-Thomas distribution $p(x)=\mathcal N e^{-x \mathcal N}$~\cite{porter1956fluctuations}.  
 
The key objective of this paper is to understand whether $\Delta_2^{(k)}(t)$ approaches 0 at large times and, if so, how large should one take $t$ to have that $\Delta_2^{(k)}(t)$ is certainly smaller than some arbitrary $\epsilon$. When this happens we say that the distribution in Eq.~\eqref{eq:ensemble} forms a $\epsilon$-approximate $k$-design. 

As discussed in the introduction this is a highly interesting question: showing that 
\be
\label{eq:deltatozero}
\Delta_2^{(k)}(t) \to 0, \qquad \forall k, 
\ee
sufficiently fast in time, we would prove that it is possible to generate (pseudo) random states using shallow quantum circuits with only a minimal amount of randomness. Importantly Eq.~\eqref{eq:deltatozero} does not only concern averages: it is a statement about the full distribution of states generated by the circuits. This means, in particular, that all the quantities that are self-averaging according to the Haar measure can be computed using the states generated by a single realisation of the minimally random circuits, and hence with a much more limited use of resources. 

Before concluding this section we note that $\Delta_2^{(k)}(t)\leq\epsilon$ is not the only possible definition of $\epsilon$-approximate $k$-design: a different, widely used one is the one proposed in Ref.~\cite{brandao2016local}. In this paper, however, we only use $\Delta_2^{(k)}(t)\leq\epsilon$.

\section{Numerical Survey}
\label{sec:numericalsurvey}

To get some intuition on the behaviour of $\Delta_2^{(k)}(t)$ for large times we begin with a \emph{numerical survey} considering both Cases~\ref{CaseA} and~\ref{CaseB}. We focus our attention on a class of circuits with $d=2$ characterized by the following parametrization 
\begin{eqnarray}
\label{eq:staticgates}
       U_{n,n+1}^{(n)} = e^{i\left[ J_x \sigma^x\otimes \sigma^x + J_y \sigma^y\otimes \sigma^y + J \sigma^z \otimes \sigma^z \right]} u_{n} \otimes u_{n+1},
\end{eqnarray} 
where $J_x=J_y = {\pi}/{4}-\delta$ is fixed, and the one-site unitaries are characterized by fixed Euler angles
\begin{equation}
    u = \exp[i \theta_1  \sigma^z] \exp[i \theta_2  \sigma^x] \exp[i \theta_3  \sigma^z].
\end{equation}
Moreover --- unless we explicitly state otherwise --- we take our initial state to be the all zero state
\begin{equation}
  \ket{\psi} = \ket{0}^{\otimes 2L}.
\end{equation}
Throughout this section $\delta$ is varied, however, we pay special attention to the point $\delta = 0$ where the circuit is DU (cf.~Sec.~\ref{sec:DU}). In Case~\ref{CaseA} we fix $J=0.1$ and randomly generate one realisation of the random Euler angles for each $u_j$, but we keep this fixed performing no average. In Case~\ref{CaseB} the one-site unitaries $\{u_{n}\}$ are taken to be Haar random while we compare different $J$s. We discuss separately the two cases.

\subsection{Case~\ref{CaseA}}

Since we have a single random unitary matrix at the boundary of the system (at $x=2L$), in Case~\ref{CaseA} we can evaluate Eq.~\eqref{eq:2norm} in two ways: We can either compute the average explicitly and then calculate the dynamics, or numerically average over circuit realisations. Both approaches are discussed in more detail in Sec.~\ref{sec:analysis}.

\begin{figure}[t]
\centering
\includegraphics[width=0.96\linewidth]{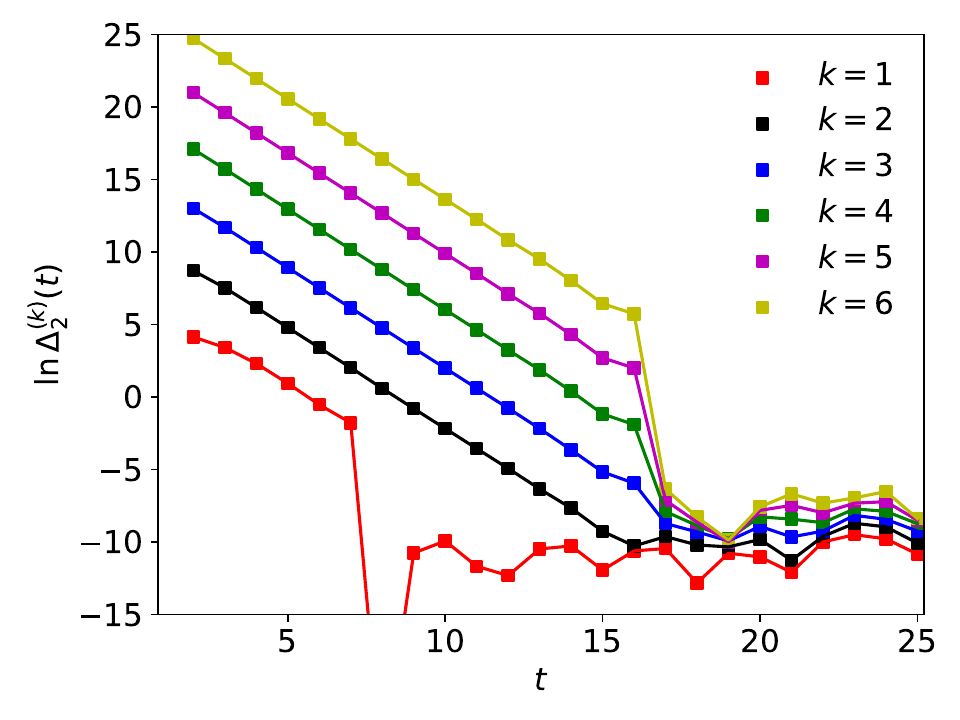} 
\includegraphics[width=0.96\linewidth]{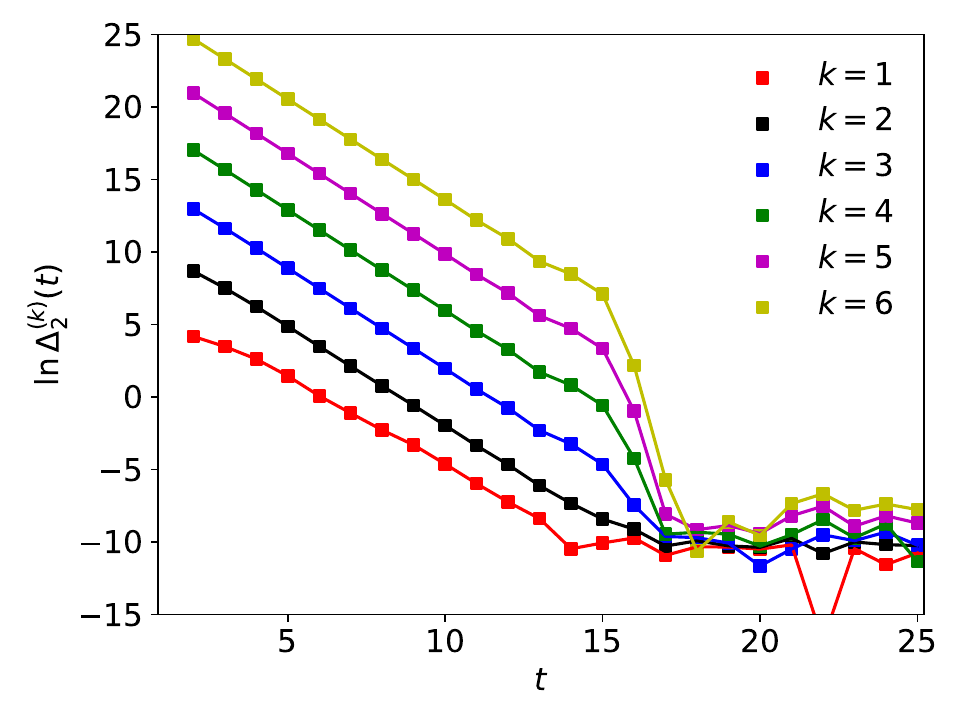} 
\caption{Calculating $\Delta_2^{(k)}(t)$ with sampled averaging for $L = 4$ (8 qubits) for various $k$. Here we take random Pauli's on the boundary and evaluate the expression in Eq. \eqref{eq:framepot} for all configurations for $t< 9$. For $t\geq 9$ we take $10^{9}$ samples. We set $\delta = 0$ in the top figure and $\delta = 0.1$ in the bottom figure.}
\label{fig:highermoms}
\end{figure}

\begin{figure*}[ht!]
\centering
\!\includegraphics[width=0.33\linewidth]{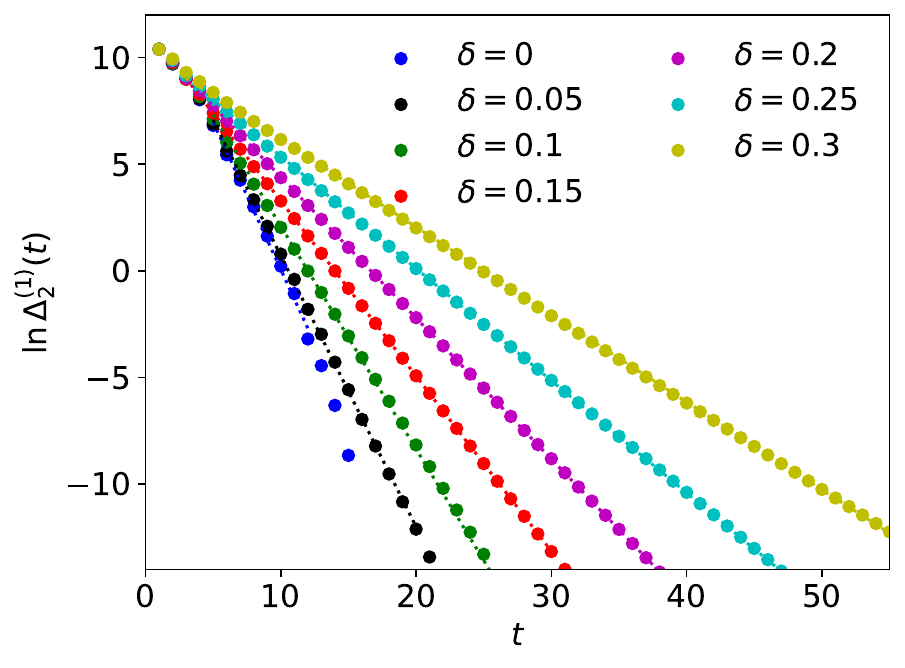}\!
\includegraphics[width=0.33\linewidth]{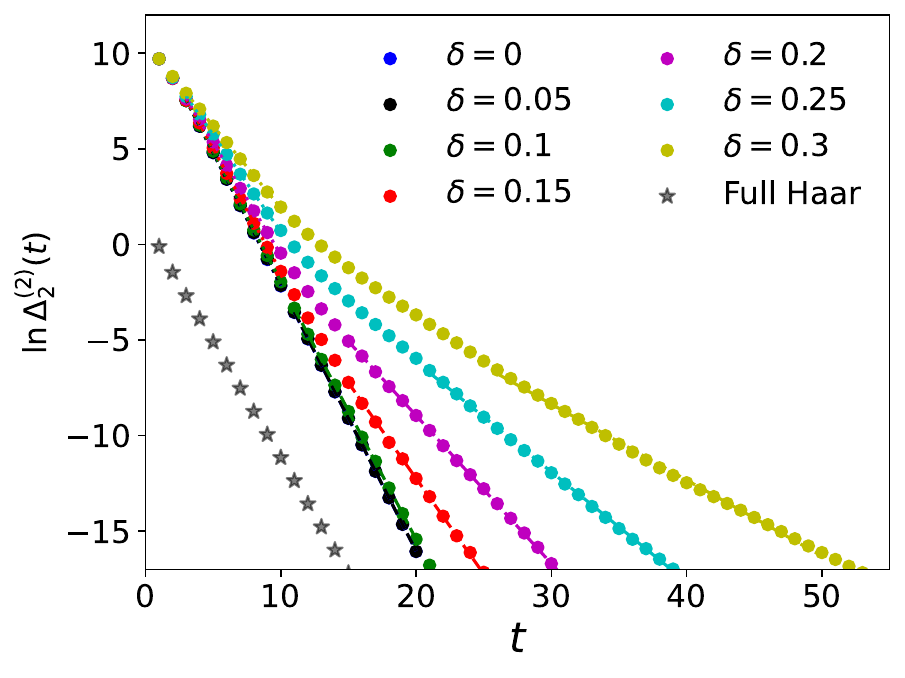} \!
\includegraphics[width=0.33\linewidth]{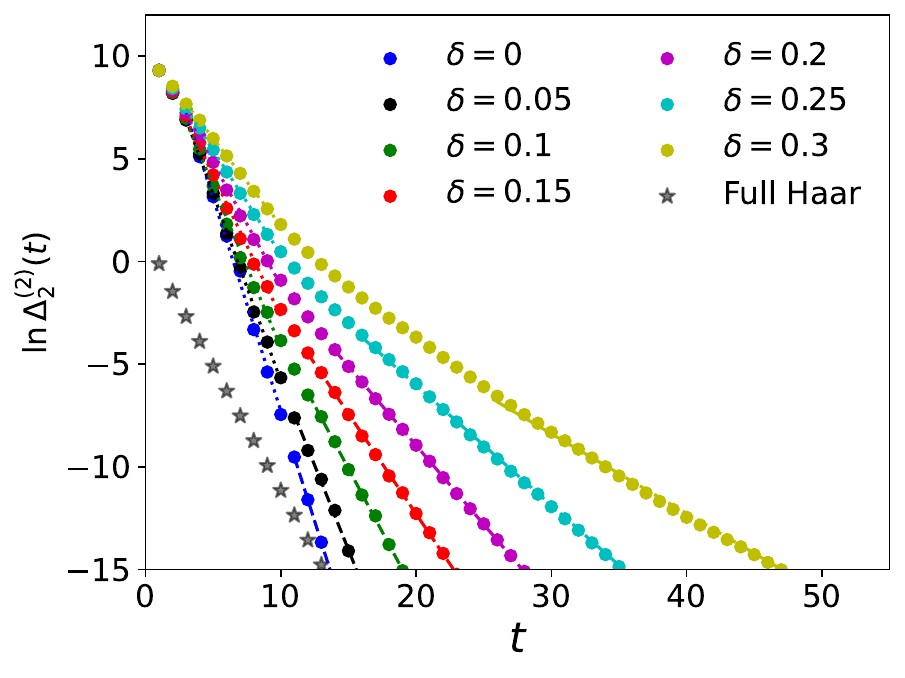}
  \caption{Calculating $\Delta_2^{(k)}(t)$ with explicit averaging for various $\delta$ and $k=1,2$. Data for $k=1$ is from system size $L=8$ (16 qubits). (Left) curve captures both random Pauli and Haar averaging on the boundary for $k=1$. Data for $k=2$ is from system size $L = 4$ (8 qubits). (Middle) Boundary randomness is random Pauli matrices [Case~\ref{CaseA1}] and $k=2$. (Right) Boundary randomness is with one site Haar random unitaries [Case~\ref{CaseA2}] and $k=2$.}
\label{fig:explicitk2}
\end{figure*}

First we perform the average numerically in the case of random Pauli matrices at the boundary, i.e., considering the choice~\ref{CaseA1} in the language of the previous section. We approximate the average by sampling a maximum of $10^{9}$ configurations of Paulis in Eq~\eqref{eq:framepot}. This method allows us to compute any moment $k$ exactly for sufficiently small times ($t\in [0,8]$) while larger values of $t$ may be approximated. A representative example of our results is presented in the top panel of Fig.~\ref{fig:highermoms}, which displays dynamics of the moments $k\in[1,6]$ at the DU point $(\delta=0)$ and a non-DU point $\delta = 0.1$. We see that the distance seems to decay quickly in time, suggesting that the minimally random circuits do produce $k$-designs at large times --- this statement can in fact be proved exactly as discussed in Sec.~\ref{sec:analysis}. We also remark that our approximation scheme succumbs to noise soon after $t=13$. Indeed, the approximate plateau shown in the figure is a  breakdown in the approximation, as we will see in Fig. \ref{fig:explicitk2} by computing $\ln \Delta_2^{(k)}(t)$ via a different method. 

Analysing the results more closely we see that, for ${k>1}$, the DU point shows a common slope of
\be
\label{eq:earlytimeDelta}
\ln \Delta_2^{(k)}(t) \approx -r_1^{(k)} t + c_1^{(k)},
\ee
where $r_1^{(k)} = \ln 4 \approx 1.39$ and $c_1^{(k)} \sim k L\ln 2$. The latter value is found by applying the Stirling approximation to Eq.~\eqref{eq:Haarmoments}, and holds regardless of $\delta$ or the specific form of the initial state (as long as it is pure). Instead, for ${k=1}$ the state $\rho_t^{(k)}$ reaches exactly the maximally mixed state --- first moment of the Haar ensemble ---  in a finite number of steps equal to $2L+1$. As we show in Sec.~\ref{sec:DU}, this is a direct consequence of dual-unitarity and it is well captured by the numerical average even at large times, when $4^{2t} \gg 10^9$. 
 
In the bottom figure of Fig.~\ref{fig:highermoms} we see the dynamics for a non-DU point with $\delta = 0.1$. Here the slope is again similar for all $k$ but slower than the DU point with $r_1^{(k)} \approx 1.36$. A slowing down of the decay when departing from the dual-unitary point can be expected by recalling that DU circuits produce a maximally fast growth of entanglement and operator spreading (see Ref.~\cite{bertini2025exactly} and references therein). Ultimately, however, this behaviour is far from obvious as $\Delta_2^{(k)}(t)$ is not an entanglement related quantity.

Sampling the configurations of $\boldsymbol{\alpha}_t$ numerically becomes unfeasible for larger times. To progress, we evaluate the average in Eq.~\eqref{eq:kmoment} explicitly, coupling together $2k$ copies of the system. In this way the evolution of $\rho_t^{(k)}$ is obtained by numerically constructing the superoperator $\mathcal B_k[\cdot]$ generating its dynamics (cf.\ Sec.~\ref{sec:analysis}), i.e.
\be
\rho_{t+1}^{(k)} = \mathcal B_k[\rho_t^{(k)}]\,.
\ee
The operator $\mathcal B_k$ inherits the brickwork structure from $\mathbb U$, but the local space of its gates is of dimension $d^{2k}$, rather than $d$. This essentially limits our investigation to $k=1,2$ (for Case~\ref{CaseB} also $k=3$ is achievable) but allows us to explore an essentially \emph{arbitrary} time window for moderate system sizes.

The possibility of exploring large times reveals an interesting phenomenon. These dynamics are reported in Fig. \ref{fig:explicitk2}. While away from the DU point $ \Delta_2^{(k=1)}(t)$ shows a convincing exponential relaxation characterised by a single rate $r^{(1)}$, $ \Delta_2^{(k=2)}(t)$ shows \emph{two-step relaxation} of the form
\begin{equation} \label{eq:twostep}
    \ln \Delta_2^{(2)}(t) \simeq \begin{cases}
        -r_1^{(2)} t + c_1^{(2)}, & t\leq t_*^{(2)} \\
        -r_2^{(2)} t + c_2^{(2)} & t> t_*^{(2)}
    \end{cases},
\end{equation} 
where $t_*^{(2)}$ is generically $\delta$-dependent. A similar phenomenon was recently observed in the dynamics of purity~\cite{bensa2021fastest, znidaric2022solvable, znidaric2023phantom} and out-of-time-ordered correlators~\cite{bensa2022two, znidaric2023two} in quantum circuits of finite size. First, in the case structured random circuits like those of our Case~\ref{CaseB}~\cite{bensa2021fastest, znidaric2022solvable, znidaric2023phantom, bensa2022two}, and then also in non-random cases~\cite{znidaric2023two}. This behaviour was explained in Ref.~\cite{Jonay2024} (see also Ref.~\cite{jonay2024twostage}) through a mapping to a statistical mechanical model (we comment more on this in Sec.~\ref{sec:DU}). 

\begin{figure}[h!]
\centering
\includegraphics[width=0.96\linewidth]{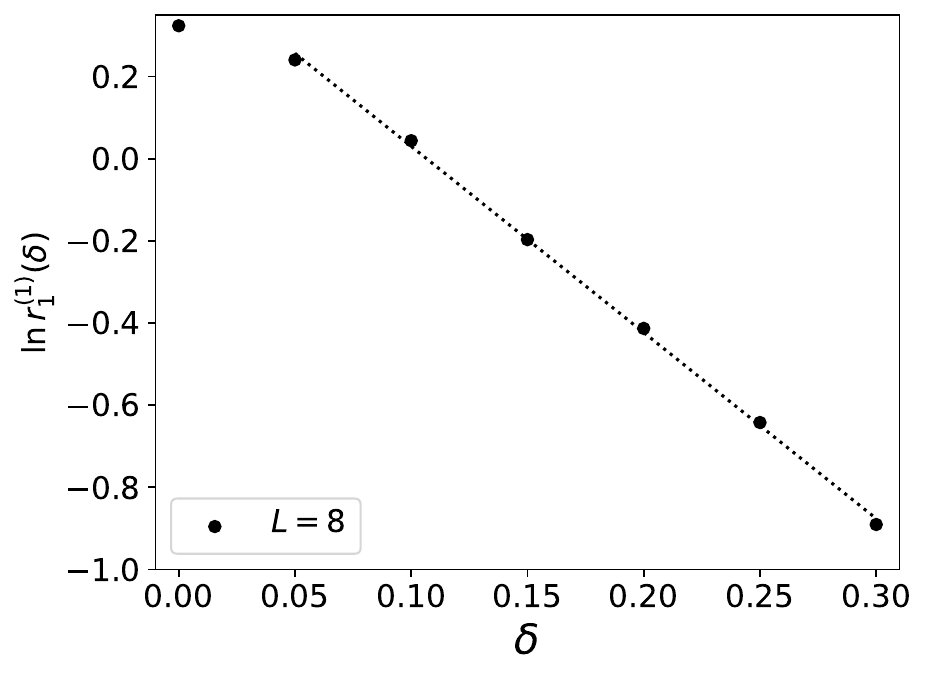}
\caption{$r_1^{(1)}$ for various $\delta$ extracted from the $L = 8$ (16 qubits) data presented in the left plot of Fig. \ref{fig:explicitk2}.}
\label{fig:r1r2k1}
\end{figure}

Consider first the results for $k=1$ (left panel of Fig.~\ref{fig:explicitk2}). Note that these correspond to both choices~\ref{CaseA1} and~\ref{CaseA2} because for ${k=1}$ the average over random Paulis and one-site random unitary matrices coincide. We already observed that in the DU case we have $\Delta_2^{(1)}(t\geq 2L) = 0$, while  
\begin{equation}
  \left.r^{(1)}\right|_{\substack{\delta=0\\ t<2L}}
  =\left.\ln\frac{\Delta_2^{(1)}(t-1)}{\Delta_2^{(1)}(t)}\right|_{\delta=0}
  =\ln d^2=\ln 4.
\end{equation}
Increasing $\delta$, however, we observe a convincingly exponential decay after a few time steps for all $\delta>0$. In particular, we have $r^{(1)}(\delta) \leq \ln 4$, confirming our previous observation that even before relaxation the decay \emph{is fastest at the DU point} and it is monotonically slower away from it. We demonstrate this in Fig.~\ref{fig:r1r2k1}, where, for ${\delta>0}$, $r_1^{(1)}$ shows a roughly exponential dependence on $\delta$ i.e.,  
\begin{equation}
    r_1^{(1)}(\delta) \approx \exp \left( -4.53 \delta + c \right).
\end{equation}

\begin{figure}[ht!]
\centering
\includegraphics[width=0.96\linewidth]{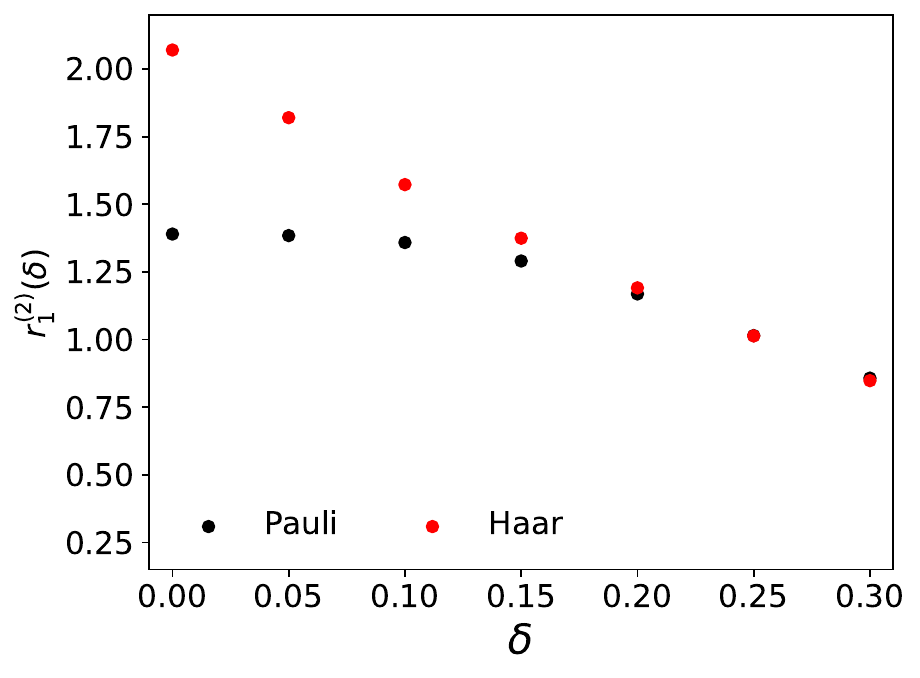} \\
\includegraphics[width=0.96\linewidth]{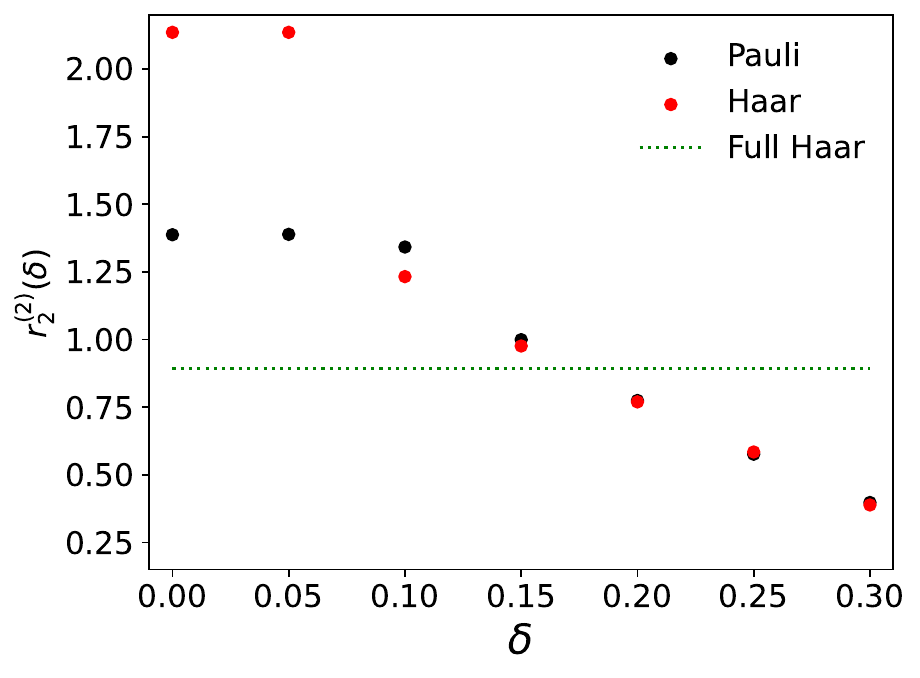} 
\caption{ $r_1^{(2)},r_2^{(2)}$ for various $\delta$ from the $L = 4$ data presented in Fig.~\ref{fig:explicitk2}. The full Haar value is calculated for $d=2$ in Eq.~\eqref{eq:haardecay}.}
\label{fig:r1r2k2}
\end{figure}

Next, we discuss our findings for $k=2$, where the Cases~\ref{CaseA1} and~\ref{CaseA2} no longer result in the same dynamics. We see this in the central and right panels of Fig.~\ref{fig:explicitk2}, which provide a comparison between the two. First we note that in this case the DU point does not show exact relaxation to the Haar value in a finite number of time-steps, however, it asymptotically flows to it faster than $\delta>0$. We see a two step relaxation for both boundary drivings and all ${\delta>0}$, while the DU point has an approximately flat decay with ${r_1^{(2)} = r_2^{(2)} \sim  \ln d^2 = \ln 4}$ for Pauli driving and ${r_1^{(2)} = r_2^{(2)} \sim \ln d^3 = \ln 8}$ for Haar driving. In Fig.~\ref{fig:r1r2k2} we plot our extracted $r_1^{(2)}, r_2^{(2)}$ for both the cases studied. Close to the DU point the Haar driving is significantly faster than Pauli, while for sufficiently large $\delta$ the decay rates become equivalent. For both cases, all instances with $\delta <0.15$ generate decay rates faster than that of the Haar random brickwork circuit in the limit of large size (cf.~\cite{brandao2016local, brandao2016efficient, hunterjones2019unitary, suzuki2024global}), i.e., 
\begin{equation}
\label{eq:haardecay}
    r_2^{(2)}\big|_{\rm Haar} = 4 \ln \frac{d^2+1}{2d}. 
\end{equation} 
We also generically observe $r_1^{(2)} > r_2^{(2)}$.

\begin{figure}[ht!]
\centering
\includegraphics[width=0.96\linewidth]{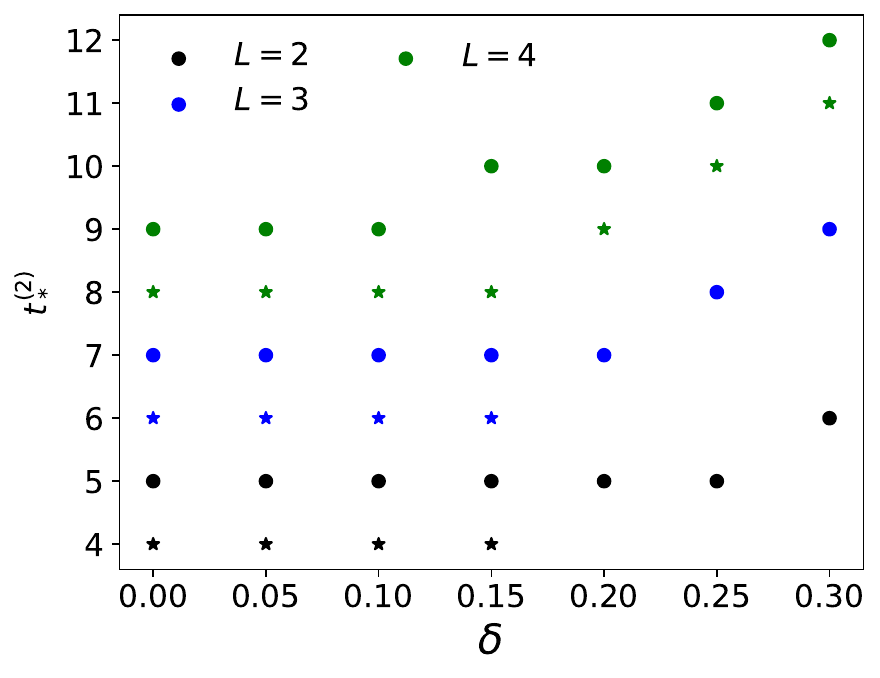} \\
  \caption{$t_*^{(2)}$ as a function of $\delta$, extracted from data given in the figures in Fig. \ref{fig:explicitk2} and similar data sets for $L=2,3$. $t_*^{(2)}$ is extracted as the first point incompatible with the first decay regime. Solid circle are from Case~\ref{CaseA1} and stars are Case~\ref{CaseA2}.}
\label{fig:tstartk2}
\end{figure}

Finally, Fig.~\ref{fig:tstartk2} reports the transition time from the first to the second decay regime. We see weak dependence on $\delta$ when the latter is small, while the second decay regime takes over at later times for larger $\delta$. The $L$ dependence close to the DU point is consistent with ${t_*^{(2)} = 2L}$ for Haar driving and ${t_*^{(2)} = 2L+1}$ for Pauli driving. Larger $\delta$s appear to pick up a constant shift, so that for all $\delta$s the scaling behaviour is consistent with $t_*^{(2)} \sim 2L$ for large $L$.

\subsection{Case~\ref{CaseB}}
We now move to investigate an example of Case~\ref{CaseB}, i.e., random onsite matrices everywhere. Specifically, Case~\ref{CaseB1} in the classification of Sec.~\ref{sec:setting}. Here we have Haar random unitary driving on every qudit for each time step but the two qudit interaction is fixed. The details on the specific implementation of this setting are presented in Sec.~\ref{sec:analysis}. For this class of circuits (and small $k$) we can study much larger systems due to the effective reduction in local dimension provided by the average over random one-site matrices. In light of the similar behaviours observed in the purity's evolution (cf.\ Ref.~\cite{bensa2021fastest,znidaric2022solvable,znidaric2023phantom}), we expect Case~\ref{CaseB} to show similar phenomenology to Case~\ref{CaseA}. 

\begin{figure}[h]
\centering
\includegraphics[width=0.96\linewidth]{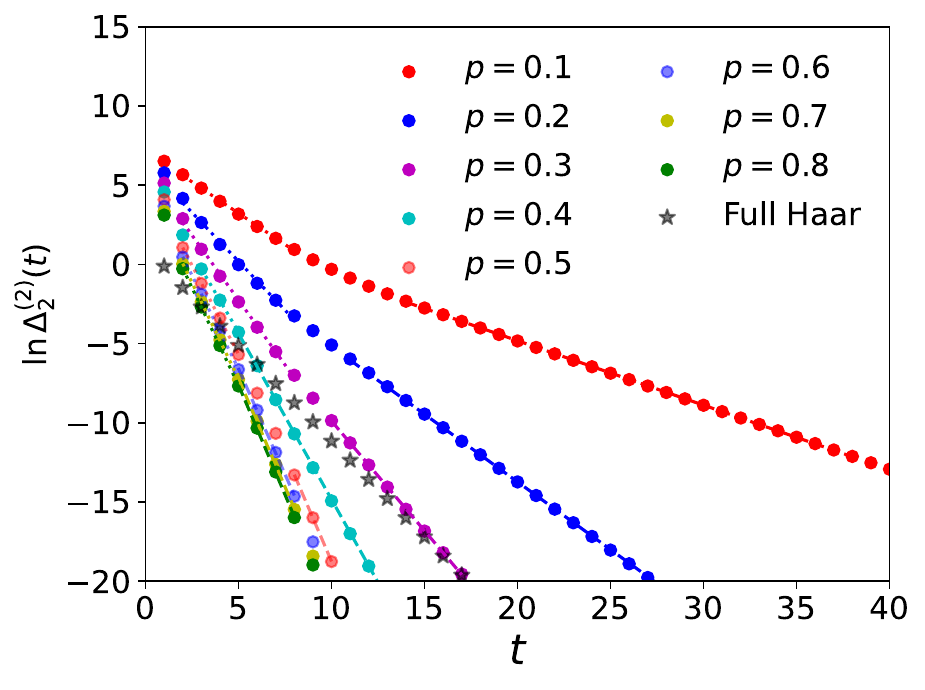} \\
\includegraphics[width=0.96\linewidth]{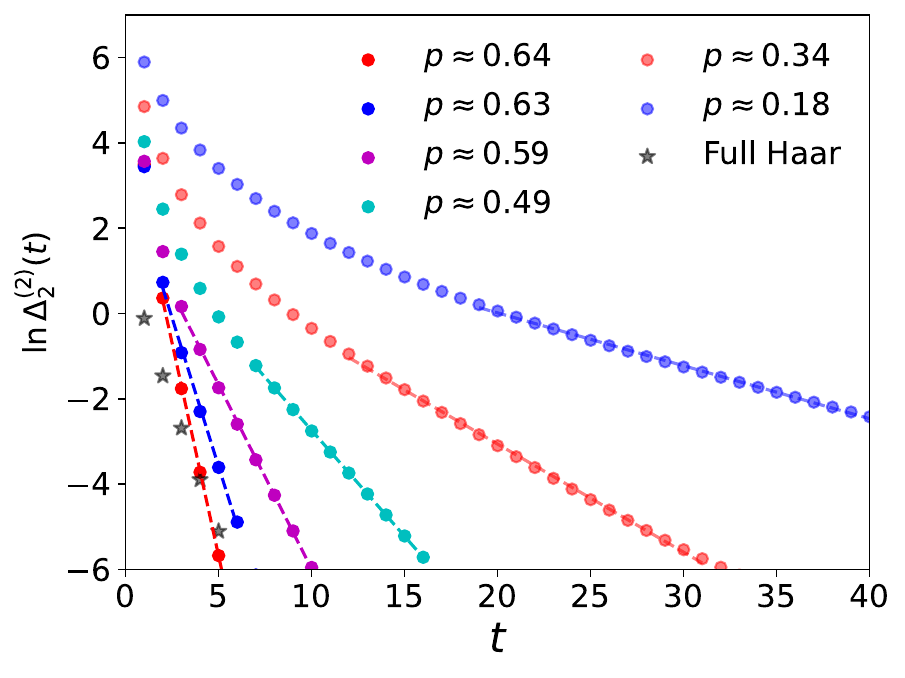}
  \caption{$\Delta_2^{(2)}(t)$ computed with explicit averaging over Case~\ref{CaseB1} circuits. (Top) Dynamics generated by DU gates defined in Eq. \eqref{eq:avgDU} for $L = 14$ and $k=2$. We set $d = 2$ for this example. (Bottom) Non-DU gates using the parametrization of Eq. \eqref{eq:staticgates}. We set $J=0.1$, $L=14$, and vary $\delta$. In the figure we report the entangling power $p$ of the averaged gate, which corresponds to (from high to low entangling power) an evenly spaced set of $\delta = 0.1\ldots 0.6$. These values were chosen to correspond to $\delta >0$ values matching the data from Fig.~\ref{fig:explicitk2}, as well as extending it to $\delta=0.4,0.5,0.6$ to give a variety of entangling powers. Both these cases display the two-step relaxation described in Eq.~\eqref{eq:twostep}.}
\label{fig:AVGDUdynamics}
\end{figure}

In Fig.~\ref{fig:AVGDUdynamics} we present results for the evolution of  $\Delta_2^{(2)}(t)$ for different DU gates parametrised by their entangling power (cf.\ the discussion in Sec.~\ref{sec:DUcaseB})
\begin{equation}\label{eq:defEntanglingPower}
p = \frac{2}{3} \cos (2J)^2.
\end{equation}
First we note that the two step relaxation of Eq.~\eqref{eq:twostep} is only observed for small entangling power $p<0.4$ in Fig.~\ref{fig:AVGDUdynamics}. For large $p$ we instead see a state dependent relaxation prior to settling into an exponential decay with a roughly $p$ independent rate. We report the fits for the decay rates in Tab.~\ref{tab:AVGDUr1}. For comparison, the decay rate in Eq.~\eqref{eq:haardecay} of the Haar random brickwork circuit for $d=2$ gives $r_2^{(2)} \approx 0.89257$. The small entangling power $p=0.2$ has a similar decay rate to the Haar random brickwork circuit, while the cases tested with higher entangling power $p\geq 0.3$ decay significantly faster. For comparison, we recall that the Haar circuit has an averaged entangling power $p|_{\rm Haar}=1-2/(d^2+1)$, giving $p|_{\rm Haar}=0.6$ for $d=2$. This shows that the decay speed is not only affected by the entangling power but also by the vicinity to the dual unitary point.

Next we move away from the DU point but continue to perform averaging as prescribed in Case~\ref{CaseB1}. To do this we return to the parameterisation in Eq.~ \eqref{eq:staticgates}. We fix $J=0.1$ and vary $\delta$. This is shown in the bottom plot of Fig.~\ref{fig:AVGDUdynamics}. Again we only see the two step relaxation as defined in Eq.~\eqref{eq:twostep} for small $p$ and a state dependent short time regime for larger $p$. The second decay regime is observed for all $p$, and $r_2^{(2)}$ is reported in Tab.~\ref{tab:AVGDUr1}. We note that the correlation between $p$ and the decay rate is weaker than in the DU case. For example, in Fig.~\ref{fig:AVGDUdynamics} we have two points with similar entangling power $p\approx0.64$ and $p\approx 0.63$ and visibly different decay rates. When the entangling power is similar, gates that are closer to the DU point appear to generate an appreciably faster decay (we have respectively $\delta = 0.1$ and $\delta=0.2$ in the aforementioned example). Moreover, for non-DU averaged gates we see that more entangling power is necessary to be as fast as the Haar random case result with $p\approx 0.59$ giving us a result similar to the Haar rate given in Eq.~\eqref{eq:haardecay}. For higher entangling power, however, we observe rates significantly faster than Haar random. 

\begin{table}[h]
  \begin{tabular}[t]{ccc}
    DU  & $\qquad$ & Non-DU \\
    \begin{tabular}[t]{|c||c|}
      \hline 
      $p$ & $r_2^{(2)}$ \\ \hline \hline
      $0.1$ & $0.41$ \\ \hline
      $0.2$ & $0.85$ \\ \hline
      $0.3$ & $1.39$ \\ \hline
      $0.4$ & $2.10$ \\ \hline
      $0.5$ & $2.74$ \\ \hline
      $0.6$ & $2.77$ \\ \hline
      $0.7$ & $2.73$ \\ \hline
      $0.8$ & $2.77$ \\ \hline
    \end{tabular} & & 
    \begin{tabular}[t]{|c||c|}
      \hline
      $p$ & $r_2^{(2)}$  \\ \hline\hline
      $0.18$ & $0.12$ \\ \hline
      $0.34$ & $0.25$ \\ \hline
      $0.49$ & $0.50$ \\ \hline
      $0.59$ & $0.86$ \\ \hline
      $0.63$ & $1.40$ \\ \hline
      $0.64$ & $2.00$  \\ \hline
    \end{tabular}
  \end{tabular}
    \caption{Decay rates for different entangling power $p$, extracted from both plots in Fig. \ref{fig:AVGDUdynamics} in Case~\ref{CaseB1}.}
    \label{tab:AVGDUr1}
\end{table}

\begin{figure}[h]
\centering
\includegraphics[width=0.96\linewidth]{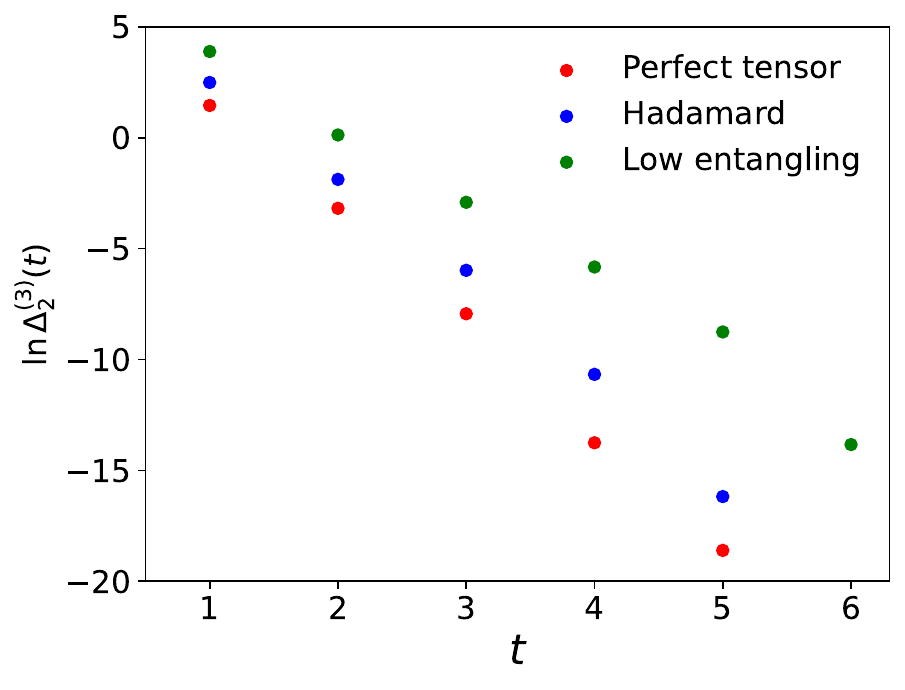}
  \caption{Time-dependence of $\Delta_2^{(3)}(t)$ for three instances of Case \ref{CaseB1}. The non-random 2-qudit gate in Eq.~\eqref{eq:refBdeftwositeU} is taken to be either a perfect tensor, cf.\ Eqs.~(\ref{eq:unitarityfoldedk}--\ref{eq:PTk}) (red), a Hadamard gate $U^{ab}_{cd}=\delta_{a,d}\delta_{b,c} \exp(\mathrm{i} 2\pi a b/d)$ (blue), or a dual-unitary gate of the form $U^{ab}_{cd}=\delta_{a,d}\delta_{b,c} \exp(\mathrm{i} J_{ab})$ (green). In the simulations we set ${d=3}$, and chose $(J_{00},J_{01},\ldots,J_{22})=(6.03, 4.24, 2.86, 5.68, 5.84, 1.34, 4.44, 3.51, 2.62)$.}
\label{fig:k3dynamics}
\end{figure}

Finally, we also consider $k=3$. Specifically, in Fig. ~\ref{fig:k3dynamics} we report the dynamics of $\Delta_2^{(k=3)}(t)$ for three static gate choices of DU gates with $d=3$. In this case, after averaging, our local Hilbert space has dimension $6$, limiting to $2L=10$ the maximal system size that we can reach. Due to this constraint we only witness early time state-dependent dynamics, however, we see quick convergence of the $3$-moments towards the Haar value. The rate at which the dynamics approach the Haar distribution is again governed by the entangling power, and it is maximal for \emph{perfect tensors}~\cite{pastawski2015holographic, helwig2012absolute, helwig2013absolutelymaximallyentangledqudit, facchi2008maximally, goyeneche2015absolutely}. The latter can be defined as the family of dual-unitary gates with optimal entangling power ($p=1$) and are discussed further in Sec.~\ref{sec:DUkg2}.

\section{Analysis}
\label{sec:analysis}

Let us now provide an analytical characterisation of the behaviours discussed in the previous section. We begin by introducing a theoretical framework to approach the problem and discuss some general features --- including the rigorous statement on the production of quantum state $k$-designs at sufficiently large times. We then specialise the treatment to the case where the local gates in Eq.~\eqref{eq:circuit} are DU, this allows us to obtain further analytical results supporting our interpretations.

Since the settings of interest in this work (cf.\ Sec.~\ref{sec:setting}) involve quantum circuits with randomness without time correlations, the time evolution of $\rho_t^{(k)}$ (cf.~Eq.~\eqref{eq:kmoment}) is \emph{Markovian}, i.e., the $k$-moment at time $t+1$ only depends on that at time $t$. More precisely we have
\be
\rho_{t+1}^{(k)} = \mathcal B_k[\rho_{t}^{(k)} ],
\ee
where we introduced the linear super operator $\mathcal B_k$: ${\rm End}(\mathbb C^{d^{2kL}})\rightarrow {\rm End}(\mathbb C^{d^{2kL}})$ defined by 
\begin{equation} \label{eq:markovop}
    \mathcal B_k[A] = 
    \smashoperator{\sum_{\vec\alpha_t \in S}}
    \mathcal{U}(\vec\alpha_t)^{\otimes k} A\, \mathcal{U}^\dag(\vec\alpha_t)^{\otimes k} \mu_S({\vec{\alpha}_t})\,,
\end{equation}
where $A\in {\rm End}(\mathbb C^{d^{2kL}})$. As it is apparent from its definition --- $\mathcal B_k[\cdot]$ is written in Krauss form with unitary Krauss operators --- this super operator is a unital and trace preserving quantum channel, i.e., a unital CPTP map~\cite{bengtsson2007geometry}. Moreover, it is additionally non-expanding, i.e., its spectrum lies on the unit disk of the complex plane.

An analysis of the fixed points of this channel allows us to answer a basic question about $k$-designs generation in the settings under exam. Indeed, straightforwardly adapting a theorem by Ippoliti and Ho (cf.\ Sec.~3~C of Ref.~\cite{Ippoliti2023}), we can show that for almost all choices of two-site gates, the brickwork circuit in Eq.~\eqref{eq:circuit} does indeed produce quantum state $k$-designs for sufficiently large times. More precisely, denoting by $\mathcal A_V\subset U(d)$ the open set containing $V\in U(d)$, we have the following rigorous statement 
\begin{property}
\label{prop:p1}
  For all the three Cases~\ref{CaseA1}, \ref{CaseA2}, and~\ref{CaseB1} the limit state  
\begin{equation}
  \lim_{t\to\infty} \rho_t^{(k)} = \rho^{(k)}_\infty,
\end{equation}
exists and --- for almost all $\{U^{(n)}\}\in \mathcal A_V$  (cf.~Eq.~\eqref{eq:circuit}) and any $(\mathcal A_V,V)$ --- $\rho^{(k)}_\infty=\rho^{(k)}_H$.  
\end{property}
This statement guarantees that in all three Cases~\ref{CaseA1}, \ref{CaseA2}, and~\ref{CaseB1}, choosing $\{U^{(n)}\}$ in the arbitrary open set $\mathcal A_V$, we almost always produce a circuit that generates the Haar distribution at large enough times, confirming our numerical observations of the previous section. Note that, crucially, no averaging is performed over the $\{U^{(n)}\}$. See App.~\ref{app:IppolitiHotheorem} for a proof of this statement.

\begin{table*}[t]
\centering
    \begin{tabular}{|l || l || l|}
      \hline
      Setting & Result &  Numerical test  
      \\ \hline\hline Early-time, Case~\ref{CaseA}, DU, any $k$ & Eq.~\eqref{eq:decayrates}  & Figs.~\ref{fig:highermoms},\ref{fig:explicitk2},\ref{fig:r1r2k1},\ref{fig:r1r2k2} 
    \\
     \hline
        Early-time, Case~\ref{CaseB}, DU, $k=2$ & Eq.~\eqref{eq:DUearlydecay} & Fig.~\ref{fig:AVGDUdynamics}  
       \\
       \hline
       Late-time, Case~\ref{CaseB}, DU, $k=2$ & Eq.~\eqref{eq:Deltascalinglatetimek2} & Fig.~\ref{fig:AVGDUdynamics} 
       \\
        \hline
        Late-time, Case~\ref{CaseB}, highly entangling DU, $k\geq 2$ &  Eq.~\eqref{eq:Deltascalinglatetimekg2} & Fig.~\ref{fig:k3dynamics}\\
        \hline
    \end{tabular}
    \caption{Summary and comparison of the results of Secs.~\ref{sec:numericalsurvey} and \ref{sec:analysis}.}
    \label{tab:connect34}
\end{table*}

Our second question --- concerning the way in which the Haar distribution is approached --- is instead more difficult to answer. To make some progress we apply the standard vectorisation (folding) mapping
\begin{equation}
\ketbra{i}{j}\mapsto \ket{i,j},
\end{equation}
where $\{\ket{j}, j=1,\ldots, d \}$ is the local computational basis. Then we can represent $\mathcal B_k[\cdot]$ as a matrix in ${\rm End}(\mathbb C^{d^{4kL}})$. For brevity, we make a little abuse of notation and denote the matrix corresponding to $\mathcal B_k[\cdot]$ with the same symbol but remove the explicit dependence on the input, i.e., we write it as $\mathcal B_k$. The matrix $\mathcal B_k$ inherits the spectral properties of the channel $\mathcal B_k[\cdot]$. Namely, labelling its eigenvalues as $\{\lambda_{m,q}\}$ we have $|\lambda_{m,q}|\leq 1$. In this labelling $m$ ranks the eigenvalues magnitude in descending order, $|\lambda_{m,q}| > |\lambda_{m+1,q}|$, and $q$ distinguishes eigenvalues with degenerate magnitudes, $|\lambda_{m,q+1}|=|\lambda_{m,q}|\equiv|\lambda_m|$~\footnote{We use $|\lambda_m|$ without an additional subscript whenever we want to refer to the magnitude of the eigenvalue.}. Instead, we represent the vectorised form of $\rho_{t}^{(k)}$ as $\ket{\smash{\rho_{t}^{(k)}}}$. Note that 
\begin{equation}
  \ket{\smash{\rho_{0}^{(k)}}}= (\ket{\psi}\otimes\ket{\psi}^*)^{\otimes k},
\end{equation}
where $\ket{\psi}$ is the initial state (cf.\ Sec.~\ref{sec:setting}).

Putting everything together, we can represent the vectorised $k$-moment at time $t$ as
\begin{equation} \label{eq:vecevol}
    \ket{\smash{\rho_{t}^{(k)}}} 
    = \mathcal{B}_k^t \ket{\smash{\rho_{0}^{(k)}}} 
    =  \sum_{m\ge0} \sum_{q=0}^{p_m-1} \lambda_{m,q}^t c^{(l)}_{m,q} 
    \ket{\smash{\lambda^{(r)}_{m,q}}},
\end{equation}
where $\ket{\smash{\lambda^{(r)}_{m,q}}}(\bra{\smash{\lambda^{(l)}_{m,q}}})$ is the right (left) eigenvector associated to $\lambda_{m,q}$,  $p_m$ denotes the degeneracy of $|\lambda_{m,q}|$, and we set
\be
{c^{(l)}_{m,q}=\braket{\smash{\lambda^{(l)}_{m,q}}}{\smash{\rho_{0}^{(k)}}}},\quad c^{(r)}_{m,q}=\braket{\smash{\rho_{0}^{(k)}}}{\smash{\lambda^{(r)}_{m,q}}}.
\ee

Property~\ref{prop:p1} ensures that the eigenspace associated with the maximal-magnitude eigenvalue $\lambda_{0}=1$ produces the vectorised form of the $k$-th moment of the Haar ensemble in all the three cases considered here~\ref{CaseA1}. More precisely
\be
\ket{\smash{\rho_{H}^{(k)}}}  = \sum_{q=0}^{p_0-1} c^{(l)}_{0,q} \ket{\smash{\lambda^{(r)}_{0,q}}}, \quad \bra{\smash{\rho_{H}^{(k)}}} = \sum_{q=0}^{p_m-1} 
 c^{(r)}_{0,q} \bra{\smash{\lambda^{(l)}_{0,q}}}.
\ee
Therefore, Eq.~\eqref{eq:vecevol} can be re-written as 
\begin{equation}
    \ket{\smash{\rho_{t}^{(k)}}} - \ket{\smash{\rho_{H}^{(k)}}} =  \sum_{m=1} \sum_{q=0}^{p_m-1} \lambda_{m,q}^t c^{(l)}_{m,q} \ket{\smash{\lambda^{(r)}_{m,q}}},
\end{equation}
which gives 
\begin{equation}
  \begin{aligned}
    \label{eq:normspelledout}
    &\| \rho_t^{(k)} - \rho_H^{(k)}\|_2^2 = 
    \| \ket{\smash{\rho_{t}^{(k)}}} - \ket{\smash{\rho_{H}^{(k)}}} \|^2 \\
    & = \smashoperator[l]{\sum_{m,m' \geq 1}}
    \smashoperator[r]{\sum_{q,q'=0}^{p_m-1}}
    \lambda_{m,q}^t (\lambda^*_{m',q'})^t c_{m,q}^{(l)} 
    c_{m',q'}^{(l)\ast} \braket{\smash{\lambda^{(r)}_{m',q'}}}{\smash{\lambda^{(r)}_{m,q}}},
  \end{aligned}
\end{equation}
where we used that the Frobenius norm becomes the standard vector norm upon vectorisation.

This expression immediately implies that in the regime $t \gg L$ the time-evolution of $\| \rho_t^{(k)} - \rho_H^{(k)}\|_2^2$ --- and hence of $\Delta_2^{(k)}(t)$ in Eq.~\eqref{eq:2norm} --- is entirely determined by the magnitude of the dominant sub-leading eigenvalue of the matrix $\mathcal{B}_k$. Namely
\begin{equation} \label{eq:asysecondreg}
  \Delta_2^{(k)}(t) \simeq \frac{C}{k!} |\lambda_1|^{2t} d^{2 k L}, \qquad t\gg L \gg 1, 
\end{equation}
where we used Stirling's approximation for the Haar moments in Eq.~\eqref{eq:Haarmoments}. This means that one can obtain $\epsilon$-approximate $k$-designs in a time 
\begin{equation}
\label{eq:designtime}
\tau^{(\epsilon)}_k \simeq 
  \frac{1}{2\ln(1/ |\lambda_1|)}(2kL \ln d - \log \epsilon)\,,
\end{equation}  
where we neglected $O(L^0)$ contributions. 

To discuss the scaling of $\tau^{(\epsilon)}_k$ with $L$ one should recall that $|\lambda_1|$ bears an implicit dependence on both $k$ and $L$, i.e., $|\lambda_1| \equiv |\lambda_1(k,L)|$. This implies that,  although Property~\ref{prop:p1} guarantees that $ |\lambda_1(k,L)|<1$ for any finite $L$, the latter can in principle approach 1 for $L\to \infty$. Whenever one can show that this is not the case, i.e.,  
\be
\lim_{L \to \infty} |\lambda_1(k,L)| \leq \bar \lambda_1 <1, \qquad \forall k, 
\label{eq:assumptiongap}
\ee
one is guaranteed that the timescale in Eq.~\eqref{eq:designtime} scales as in Haar random circuits~\cite{hunterjones2019unitary, chen2024incompressibility}. 

Proving Eq.~\eqref{eq:assumptiongap} in general is very hard: here we advocate its validity in specific cases. Namely, in the upcoming subsection we prove that, for ${\lambda_1(k=1,L)=0}$ in minimally random circuits of DU gates. Moreover, we present compelling analytical and numerical arguments suggesting that for Case~\ref{CaseB} DU circuits ${\lim_{L\to\infty} |\lambda_1(k,L)|<1}$ for all $k\geq 2$, i.e., Eq.~\eqref{eq:assumptiongap} holds. Finally, we also show that sufficiently entangling dual-unitary circuits generate random states faster than Haar random circuits ($|\lambda_1|$ is smaller). These findings are summarised in Table.~\ref{tab:connect34}, where we also pair them with the corresponding numerical observations from Sec.~\ref{tab:connect34}.

\begin{figure}[h!]
\centering
\includegraphics[width=0.96\linewidth]{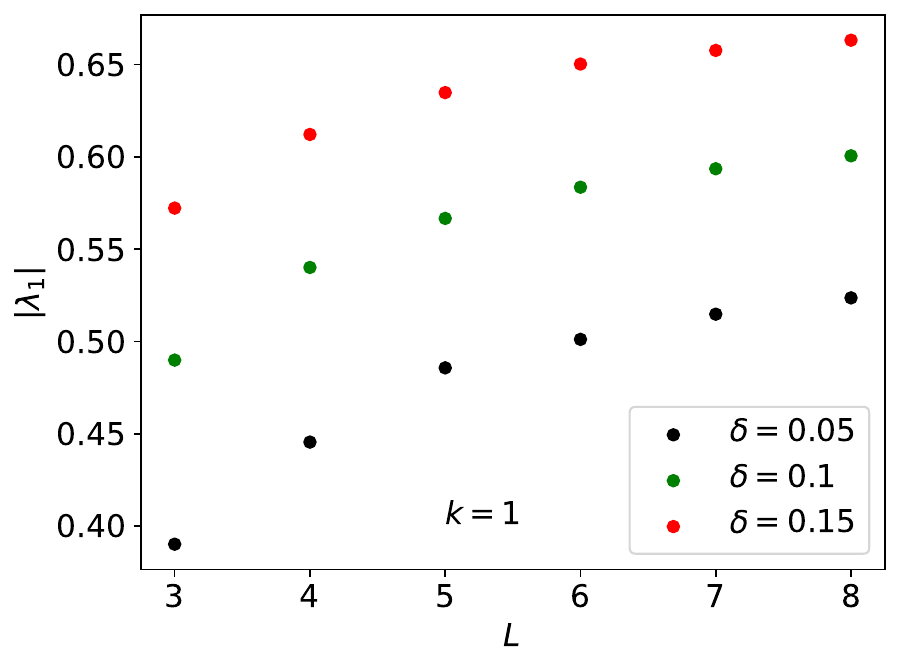}
\caption{$|\lambda_1|$ vs $L$ for for Case~\ref{CaseA} circuits with $k=1$ and different values of $\delta$. Data is extracted from  Fig.~\ref{fig:explicitk2} and similar data sets with smaller $L$.}
\label{fig:lambdak1nondu}
\end{figure}

\begin{figure}[h!]
  \centering
  \includegraphics[width=0.96\linewidth]{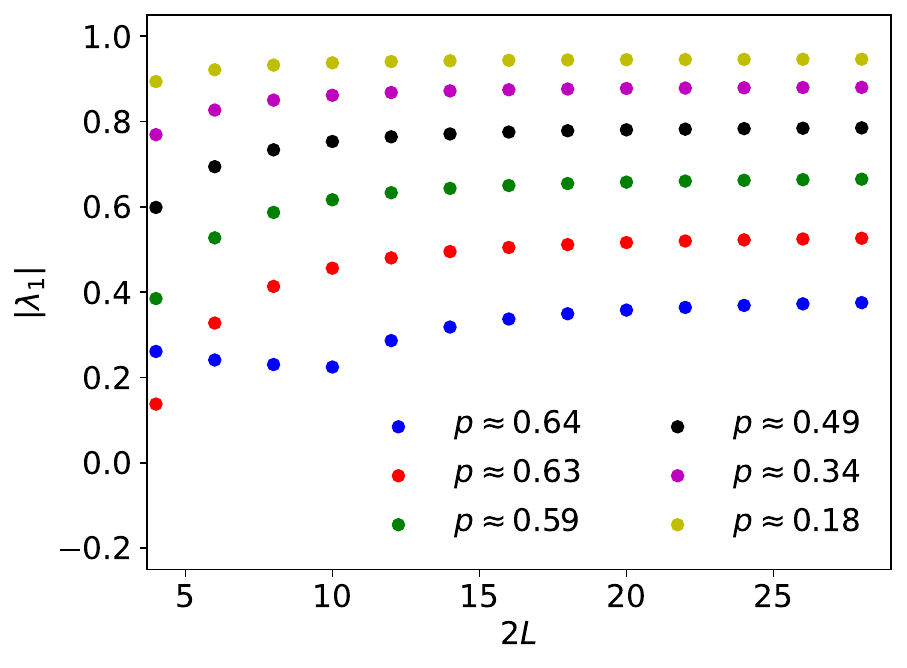}
  \caption{$|\lambda_1|$ vs $L$ for Case~\ref{CaseB} circuits with $k=2$
  away from the DU point. Data from the circuit defined for the bottom plot of Fig.~\ref{fig:AVGDUdynamics}.}
  \label{fig:casebk2nondu}
\end{figure}

\begin{figure}[ht!]
\centering
\includegraphics[width=0.96\linewidth]{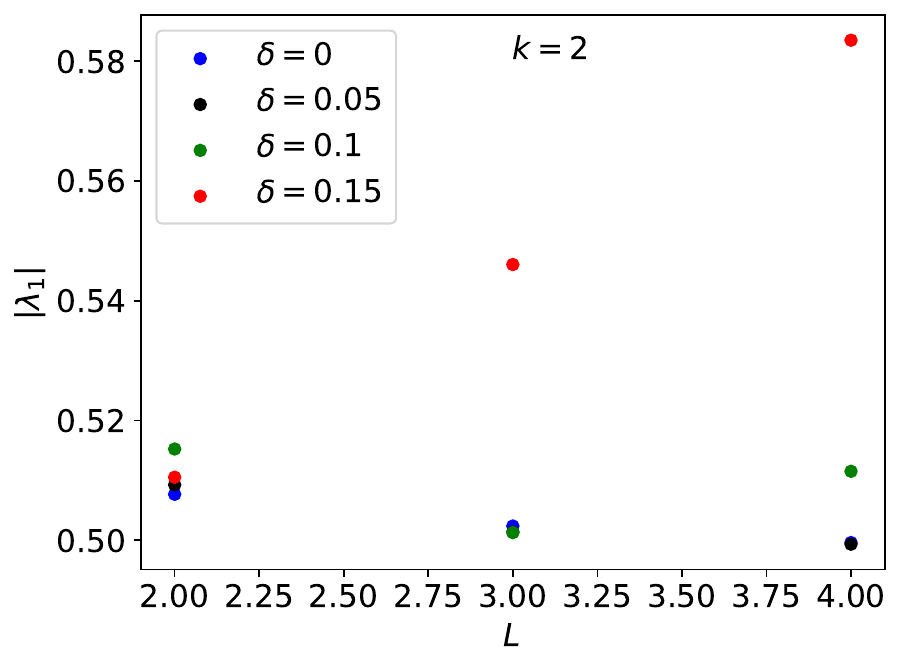} \\
\includegraphics[width=0.96\linewidth]{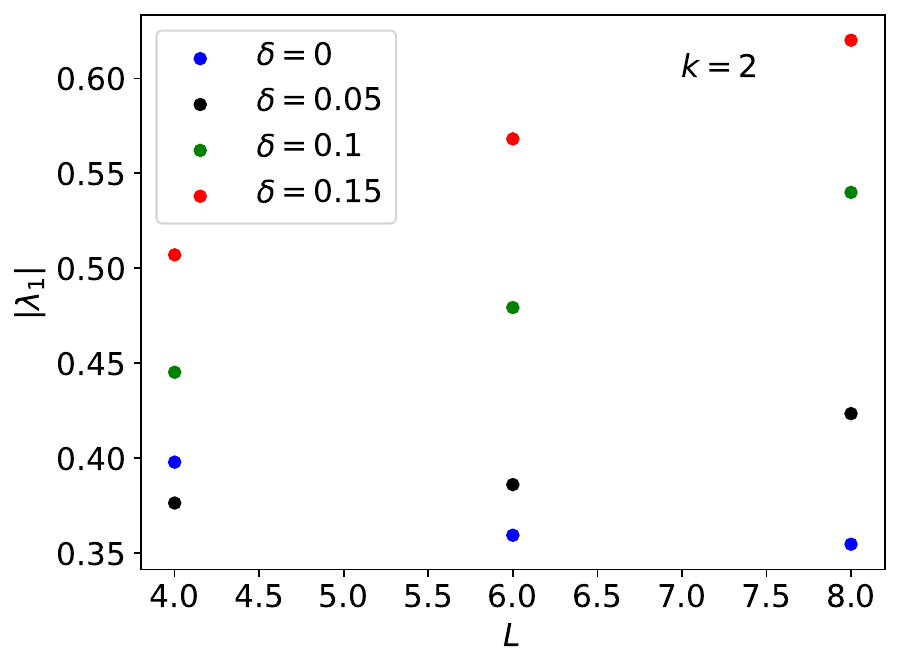} 
\caption{$|\lambda_1|$ vs $L$ for Case~\ref{CaseA} circuits  ((top) Pauli (bottom) Haar) with $k=2$ and different values of $\delta$. Data was extracted from Fig.~\ref{fig:explicitk2} and similar data sets with smaller $L$.}
\label{fig:caseak2nondu}
\end{figure}

In fact, our numerical analysis suggests that for small $k$ these statements continue to hold true also for Case~\ref{CaseA} DU circuits and for both Case~\ref{CaseA} and Case~\ref{CaseB} circuits close enough to the DU point: some representative examples of our results are reported in Figs.~\ref{fig:lambdak1nondu}, \ref{fig:casebk2nondu}, \ref{fig:caseak2nondu}. Specifically,  Fig.~\ref{fig:lambdak1nondu} considers ${\lambda_1(k=1,L)}$ vs $L$ in circuits away from the DU point ($\delta$ controls the DU breaking). The curves appear to be converging to values smaller than $1$, however, the system sizes are too small to reliably estimate this asymptotic value for $|\lambda_1|$. A similar picture is observed for $k=2$ in both Case~\ref{CaseB}  (Fig.~\ref{fig:casebk2nondu}) and Case~\ref{CaseA} (Fig.~\ref{fig:caseak2nondu}) circuits. The latter case, however, we can only access small sizes making larger values of delta hard to interpret.

Before moving to the analysis of the DU case we also emphasise that --- even when Eq.~\eqref{eq:assumptiongap} holds --- Eq.~\eqref{eq:asysecondreg} does not automatically apply in the regime $1 \ll t \ll L$, where the contribution of the higher terms in Eq.~\eqref{eq:normspelledout} is not necessarily suppressed and can change the scaling with $t$. This can lead $\Delta_2^{(k)}(t)$ to show an exponential decay with a different exponent $r_1^{(k)} \neq  2 \ln |\lambda_1|$ similarly to what observed in the other accounts of two-step relaxation~\cite{bensa2021fastest, znidaric2022solvable, znidaric2023phantom, bensa2022two, znidaric2023two, Jonay2024}. Below we show that this phenomenon occurs in the DU case.

\subsection{Dual unitary gates}
\label{sec:DU}

Let us now assume that our two-local gates are DU. Namely, we consider gates that are unitary also when propagating in space. More concretely, we consider gates for which the space-time dual matrix $\tilde{U}$ defined as 
\begin{equation}
  \mel{s_1 s_2}{\tilde{U}}{s_3 s_4}=
  \mel{s_1 s_3}{U}{s_2 s_4},
\end{equation}
is unitary, i.e. 
\begin{equation}
\label{eq:DUeq}
  \tilde{U}^{\dagger}\tilde{U}=\tilde{U}\tilde{U}^{\dagger}=\1.
\end{equation}
To aid our analytical analysis, we also consider initial states that are compatible with the dual-unitarity of the circuits, i.e., that preserve the spatial unitarity characterising DU gates. Although this choice is very beneficial for the analytical calculations, we do not expect it to affect substantially the results. We will explicitly verify this by comparing the results obtained in this section with the numerical results of Sec.~\ref{sec:numericalsurvey}, which are instead obtained for generic initial states. Specifically we consider an initial state constituted by a product of Bell pairs 
\begin{equation}\label{eq:defBellPairs}
  \ket{\psi}=\Big(\frac{1}{\sqrt{d}}\sum_{j=1}^{d} \ket{j,j}\Big)^{\otimes L},
\end{equation}
where $d$ is the local Hilbert space dimension. The reason for this choice --- which contrasts with the product states used elsewhere in this manuscript --- is that they belong to the class of \emph{solvable} initial states~\cite{bertini2019entanglement,piroli2020exact}. 

Before proceeding with a characterisation of the frame potential, we begin by introducing a convenient diagrammatic tensor-network notation, where a node with an open leg represents a vector in the space $\mathbb{C}^{d^{2k}}$, and a node with a number of outgoing legs is a vector (or a matrix/more general tensor) in the tensor product of a number of local spaces determined by the number of legs. In particular, we represent the local time-evolution operator acting on $\mathbb{C}^{d^{2k}}\otimes \mathbb{C}^{d^{2k}}$ with square boxes
\begin{equation}\label{eq:defBlueGate}
  \begin{tikzpicture}[baseline={([yshift=-0.6ex]current bounding box.center)},scale=0.5]
    \prop{0}{0}{FcolU}{k}
  \end{tikzpicture}
  = \left( U \otimes U^{\ast}\right)^{\otimes k}\mkern-24mu,\qquad
  \begin{tikzpicture}[baseline={([yshift=-0.6ex]current bounding box.center)},scale=0.5]
    \prop{0}{0}{FcolUD}{k}
  \end{tikzpicture}=
  \left.\begin{tikzpicture}[baseline={([yshift=-0.6ex]current bounding box.center)},scale=0.5]
    \prop{0}{0}{FcolU}{k}
  \end{tikzpicture}\right.^{\dagger}
  = \left( U^{\dagger} \otimes U^{T}\right)^{\otimes k}\mkern-24mu,
\end{equation}
where $\otimes$ represents the tensor product in the replicated space. Similarly, we represent with a circle the average over randomly chosen one-site unitaries
\begin{equation}\label{eq:defDk}
  \mathcal{D}_k=
  \begin{tikzpicture}[baseline={([yshift=-0.6ex]current bounding box.center)},scale=0.5]
    \RboundStr{0}{0}{colSt}{k}
  \end{tikzpicture}=
  \begin{tikzpicture}[baseline={([yshift=-0.6ex]current bounding box.center)},scale=0.5]
    \RboundStr{0}{0}{colSt}{k}
  \end{tikzpicture}^{\dagger}=\sum_{\alpha\in S} \mu_S(\alpha) 
  \left(\alpha\otimes \alpha^{\ast}\right)^{\otimes k}
\end{equation}
while the initial state $\ket*{\rho_0^{(k)}}$ is diagrammatically given as
\begin{equation}\label{eq:diagramBellPairs}
  \ket*{\rho^{(k)}_0}=
  \frac{1}{d^{k L}}
  \begin{tikzpicture}[baseline={([yshift=1.6ex]current bounding box.center)},scale=0.5]
    \def\X{10}
    \foreach \x in {2,4,...,\X}{\solstate{\x-3.5}{-0.5}}
    \draw[decorate,decoration={brace}] (\X-2.25,-0.75) -- (-1.75,-0.75) node[midway,yshift=-7.5pt] {$L$};
  \end{tikzpicture}.
\end{equation}
Using the convention that connected lines represent index summation, from the unitarity of the gates we obtain the following relation
\begin{equation}\label{eq:unitarityDiagram}
  \begin{tikzpicture}[baseline={([yshift=-0.6ex]current bounding box.center)},scale=0.5]
    \prop{0}{0}{FcolU}{k}
    \Lbound{-0.5}{0.5}
    \nRbound{0.5}{0.5}
    \prop{0}{2}{FcolUD}{k}
  \end{tikzpicture}=
  \begin{tikzpicture}[baseline={([yshift=-0.6ex]current bounding box.center)},scale=0.5]
    \Lvconn{0}{1}{-1}
    \Rvconn{1}{1}{-1}
  \end{tikzpicture}.
\end{equation}
Equivalently, from the first of Eq.~\eqref{eq:DUeq} we get
\begin{equation}\label{eq:DualUnitarityDiagram}
  \begin{tikzpicture}[baseline={([yshift=-0.6ex]current bounding box.center)},scale=0.5]
    \prop{0}{0}{FcolU}{k}
    \Lbound{-0.5}{0.5}
    \draw [thick,colLines] ({-0.5-0.25*(1+\sqrtTwo)},-0.25) -- ({-0.5-0.25*(1+\sqrtTwo)},2.25);
    \draw [thick,colLines] (-0.5,2.5) arc (45:180:0.5/\sqrtTwo);
    \draw [thick,colLines] (-0.5,-0.5) arc (135:0:-0.5/\sqrtTwo);
    \prop{0}{2}{FcolUD}{k}
  \end{tikzpicture}=
  \begin{tikzpicture}[baseline={([yshift=-0.6ex]current bounding box.center)},scale=0.5]
    \Lbound{-0.5}{0.5}
    \draw [thick,colLines] (-0.125,{-0.875+0.75/\sqrtTwo}) arc (90:45:0.75/\sqrtTwo);
    \draw [thick,colLines] (-0.125,{2.875-0.75/\sqrtTwo}) arc (90:135:-0.75/\sqrtTwo);
    \draw [thick,colLines] (-0.5,0.5) arc (45:135:-0.75/\sqrtTwo);
    \draw [thick,colLines] (-0.5,1.5) arc (135:45:0.75/\sqrtTwo);
    \draw [thick,colLines] (-0.125,{-0.875+0.75/\sqrtTwo}) arc (-90:-180:1./\sqrtTwo);
    \draw [thick,colLines] (-0.125,{2.875-0.75/\sqrtTwo}) arc (-90:0:-1./\sqrtTwo);
    \draw [thick,colLines] ({-0.125-1./\sqrtTwo},{2.875-1.75/\sqrtTwo}) -- ({-0.125-1./\sqrtTwo},{-0.875+1.75/\sqrtTwo});
  \end{tikzpicture}.
\end{equation}

\subsubsection{Case~\ref{CaseA}}

Let us now move on to consider the calculation of the frame potential $F^{(k)}_t$ in Case~\ref{CaseA}, i.e., a single one-site random matrix at the boundary. We will see that, when considering the initial states in Eq.~\eqref{eq:diagramBellPairs}, Eqs.~\eqref{eq:unitarityDiagram} and \eqref{eq:DualUnitarityDiagram} allow us to completely characterise the frame potential $F^{(k)}_t$ whenever $t<2L$.

Introducing the diagrammatic representation of the time-evolution operator $\mathcal{B}_k$, i.e., 
\begin{equation}
  \mathcal{B}_k=
  \begin{tikzpicture}[baseline={([yshift=-0.6ex]current bounding box.center)},scale=0.5]
    \def\X{10}
      \foreach \x in {4,6,...,\X}{\prop{\x-4}{0}{FcolU}{k}}
      \foreach \x in {2,4,...,\X}{\prop{\x-3}{1}{FcolU}{k}}
    \Lbound{-1.5}{-0.5}
    \Rbound{\X-2.5}{-0.5}{colSt}{k}
    \draw[decorate,decoration={brace}] (-1.75,2-0.25) -- (\X-2.25,2-0.25) node[midway,yshift=7.5pt] {$L$};
  \end{tikzpicture},
\end{equation}
where $L$ denotes the number of pairs of qudits ($L=5$ in the above example), we can express the frame potential
\begin{equation}
  F^{(k)}_t=\braket*{\rho_t^{(k)}}{\rho_t^{(k)}}=
  \mel*{\rho_0^{(k)}}{\left.\mathcal{B}_k^{\dagger}\right.^t \mathcal{B}_k^t}{\rho_0^{(k)}}
\end{equation}
as in the left diagram in Fig.~\ref{fig:FkDiagram}. The diagram is simplified by repeatedly applying Eq.~\eqref{eq:unitarityDiagram}. Proceeding in this way, and assuming $t<L$, we obtain two connected triangular tensor networks scaling with $t$, as shown in the middle panel of Fig.~\ref{fig:FkDiagram}.
\begin{figure*}
  \begin{tikzpicture}[baseline={([yshift=-0.6ex]current bounding box.center)},scale=0.5]
    \def\X{10}
    \def\Y{6}
    \node at (-2.75,\Y) {$\displaystyle \frac{1}{d^{2 k L}}$};
    \foreach \t in {2,4,...,\Y}{
      \foreach \x in {4,6,...,\X}{\prop{\x-4}{\t-2}{FcolU}{k}}
      \foreach \x in {2,4,...,\X}{\prop{\x-3}{\t-1}{FcolU}{k}}
      \foreach \x in {2,4,...,\X}{\prop{\x-3}{2*\Y-\t+1}{FcolUD}{k}}
      \foreach \x in {4,6,...,\X}{\prop{\x-4}{2*\Y-\t+2}{FcolUD}{k}}
    }
    \foreach \t in {2,4,...,\Y}{\Lbound{-1.5}{2*\Y+1.5-\t}}
    \foreach \x in {2,4,...,\X}{\Lbound{\x-3.5}{2*\Y+1.5-\Y-2}}
    \foreach \t in {2,4,...,\Y}{\Rbound{\X-2.5}{2*\Y+1.5-\t}{colSt}{k}}
    \foreach \x in {2,4,...,\X}{\nRbound{\x-2.5}{2*\Y+1.5-\Y-2}}
    \foreach \x in {2,4,...,\X}{\solstate{\x-3.5}{-0.5}}
    \foreach \x in {2,4,...,\X}{\solstateD{\x-3.5}{2*\Y+0.5}}
    \foreach \t in {2,4,...,\Y}{\Lbound{-1.5}{-2.5+\t}}
    \foreach \t in {2,4,...,\Y}{\Rbound{\X-2.5}{-2.5+\t}{colSt}{k}}
    \draw[decorate,decoration={brace}] (\X-2,\Y-0.75) -- (\X-2,-0.5) node[midway,xshift=7.5pt] {$t$};
    \draw[decorate,decoration={brace}] (-1.75,2*\Y+1-0.25) -- (\X-2.25,2*\Y+1-0.25) node[midway,yshift=7.5pt] {$L$};
    \begin{scope}[shift={({\X+4.25},{0})}]
        \node at (-2.75,\Y) {$\displaystyle \frac{1}{d^{2 k L}}$};
        \foreach \t in {2,4,...,\Y}{\Rbound{\X-2.5}{2*\Y+1.5-\t-1}{colSt}{k}}
        \foreach \x in {2,4,...,\X}{\solstate{\x-3.5}{0.5}}
        \foreach \x in {2,4,...,\X}{\solstateD{\x-3.5}{2*\Y-0.5}}
        \foreach \t in {2,4,...,\Y}{\Rbound{\X-2.5}{-2.5+\t+1}{colSt}{k}}
        \foreach \t in {1,...,\Y}{\Rvconn{\X-1.5-\t}{\Y-0.5+\t}{\Y+0.5-\t}}
        \prop{\X-3}{\Y-2}{FcolU}{k}
        \prop{\X-4}{\Y-3}{FcolU}{k}
        \prop{\X-3}{\Y-4}{FcolU}{k}
        \prop{\X-5}{\Y-4}{FcolU}{k}
        \prop{\X-6}{\Y-5}{FcolU}{k}
        \prop{\X-4}{\Y-5}{FcolU}{k}

        \prop{\X-3}{\Y+2}{FcolUD}{k}
        \prop{\X-4}{\Y+3}{FcolUD}{k}
        \prop{\X-3}{\Y+4}{FcolUD}{k}
        \prop{\X-5}{\Y+4}{FcolUD}{k}
        \prop{\X-6}{\Y+5}{FcolUD}{k}
        \prop{\X-4}{\Y+5}{FcolUD}{k}

        \Rvconn{-1.5}{2*\Y-0.5}{0.5}
        \Lvconn{-0.5}{2*\Y-0.5}{0.5}
        \Rvconn{0.5}{2*\Y-0.5}{0.5}
        \Lvconn{1.5}{2*\Y-0.5}{0.5}
        \draw[decorate,decoration={brace}] (-1.75,2*\Y-0.25) -- (1.875,2*\Y-0.25) node[midway,yshift=7.5pt] {$L-t$};
        \draw[decorate,decoration={brace}] (2.125,2*\Y-0.25) -- (\X-2.25,2*\Y-0.25) node[midway,yshift=7.5pt] {$t$};
        \draw[decorate,decoration={brace}] (\X-2,\Y) -- (\X-2,0.5) node[midway,xshift=7.5pt] {$t$};
    \end{scope}
    \begin{scope}[shift={({2*\X+2},{0})}]
      \node at (\X-6.75,\Y) {$\displaystyle \frac{1}{d^{2 k t}}$};
      \def\X{10}
      \def\Y{6}
      \foreach \t in {2,4,...,\Y}{\Rbound{\X-2.5}{2*\Y+1.5-\t-1}{colSt}{k}}
      \foreach \t in {2,4,...,\Y}{\Rbound{\X-2.5}{-2.5+\t+1}{colSt}{k}}
      \def\dif{0.5}
      \foreach \t in {2,4,...,\Y}{
        \draw [thick,colLines] ({\X-2.5},{2*\Y+1.5-\t}) arc (45:90:1./\sqrtTwo);
        \draw [thick,colLines] ({\X-2.5},{\t-1.5}) arc (-45:-90:1./\sqrtTwo);
        \draw [thick,colLines] ({\X-3-1./\sqrtTwo-\dif*(\Y-\t)},{2*\Y+1-\t}) arc (0:-90:-1./\sqrtTwo);
        \draw [thick,colLines] ({\X-3-1./\sqrtTwo-\dif*(\Y-\t)},{\t-1}) arc (0:90:-1./\sqrtTwo);
        \draw [thick,colLines] ({\X-3-1./\sqrtTwo-\dif*(\Y-\t)},{2*\Y+1-\t})-- ({\X-3-1./\sqrtTwo-\dif*(\Y-\t)},{\t-1});
        \draw [thick,colLines] ({\X-3-\dif*(\Y-\t)},{2*\Y+1-\t+1./\sqrtTwo}) -- ({\X-3},{2*\Y+1-\t+1./\sqrtTwo});
        \draw [thick,colLines] ({\X-3-\dif*(\Y-\t)},{\t-1-1./\sqrtTwo}) -- ({\X-3},{\t-1-1./\sqrtTwo});
      }
      \foreach \t in {4,6,...,\Y}{
        \draw [thick,colLines] ({\X-2.5},{2*\Y+2.5-\t}) arc (-45:-90:1./\sqrtTwo);
        \draw [thick,colLines] ({\X-2.5},{\t-2.5}) arc (45:90:1./\sqrtTwo);
        \draw [thick,colLines] ({\X-3-1./\sqrtTwo-\dif*(\Y-\t+1)},{2*\Y+3-\t-\sqrtTwo}) arc (0:-90:-1./\sqrtTwo);
        \draw [thick,colLines] ({\X-3-1./\sqrtTwo-\dif*(\Y-\t+1)},{\t-3+2./\sqrtTwo}) arc (0:90:-1./\sqrtTwo);
        \draw [thick,colLines] ({\X-3-1./\sqrtTwo-\dif*(\Y-\t+1)},{2*\Y+3-\t-\sqrtTwo}) -- ({\X-3-1./\sqrtTwo-\dif*(\Y-\t+1)},{\t-3+2./\sqrtTwo});
        \draw [thick,colLines] ({\X-3-\dif*(\Y-\t+1)},{2*\Y+3-\t-1./\sqrtTwo}) -- ({\X-3},{2*\Y+3-\t-1./\sqrtTwo});
        \draw [thick,colLines] ({\X-3-\dif*(\Y-\t+1)},{\t-3+1./\sqrtTwo}) -- ({\X-3},{\t-3+1./\sqrtTwo});
      }
      \Rvconn{\X-2.5}{\Y+0.5}{\Y-0.5}
      \draw[decorate,decoration={brace}] (\X-2,\Y) -- (\X-2,0.5) node[midway,xshift=7.5pt] {$t$};
    \end{scope}
    \draw[thick,-latex] ({3*\X-7.25},\Y) -- ({3*\X-5.5},{\Y}) node [midway,above] {DU};
    \draw[thick,-latex] ({\X-2},\Y) -- ({\X+0.5},{\Y}) node [midway,above] {unitarity};
  \end{tikzpicture}
  \caption{\label{fig:FkDiagram}Diagrammatic representation of $F^{(k)}_t$ for $t\le L$ and dual unitary gates. Using the unitarity condition~\eqref{eq:unitarityDiagram}, the diagramm simplifies into the middle panel, while the factorised result in the right-most panel follows from additional application of dual unitary condition~\eqref{eq:DualUnitarityDiagram}.}
\end{figure*}

We now apply many times the dual-unitarity condition in Eq.~\eqref{eq:DualUnitarityDiagram} to the middle-panel of Fig.~\ref{fig:FkDiagram}, to obtain the right-most panel, which reads explicitly as
\begin{equation}\label{eq:FtEarlyRegime}
  F_{t}^{(k)}=\left(d^{-2k}\tr[\mathcal{D}_k^2]\right)^t.
\end{equation}
This expression clearly depends on the choice of boundary driving confirming our observation of Sec.~\ref{sec:numericalsurvey}. In particular, evaluating Eq.~\eqref{eq:FtEarlyRegime} in Case~\ref{CaseA1}, we can see that there is no $k$-dependence
\begin{equation}
  \left.F_t^{(k)}\right|_{(a.1)}=4^{-t},
\end{equation}
while in Case~\ref{CaseA2} the frame potential reduces to the average of powers of trace of $d\times d$ random unitaries
\begin{equation}
  \left.F_t^{(k)}\right|_{(a.2)}\mkern-16mu=\left(d^{-2k}T_{d,k}\right)^t\mkern-4mu,\qquad
  T_{d,k}=\mkern-4mu\smashoperator{\int\limits_{U(d)}}\mkern-4mu dU \left|\tr[U]\right|^{2k}
  \mkern-8mu,
\end{equation}
which for generic $d$ and $k$ takes the following form~\cite{koestenberger2021weingartencalculus}
\begin{equation}
  T_{d,k} = k!^2\sum_{l=1}^{d}
  \smashoperator[r]{\sum_{\substack{s_1+\cdots+s_l=k\\s_1\ge s_2 \ge \cdots \ge s_l}}}
    \frac{\displaystyle\prod_{i=1}^{l-1}\prod_{j=i+1}^l (s_i-s_j-i+j)^2}{\displaystyle\prod_{i=1}^l (l-i+s_i)!^2}.
\end{equation}
Note that in the case of $d=2$, the above expression reduces to the $k$-th Catalan number $C_k$,
\begin{equation}
  T_{2,k}=\frac{1}{k}\binom{2k}{k-1}\equiv C_k,
\end{equation}
while for $d>k$ we have~\cite{koestenberger2021weingartencalculus},
\begin{equation}
  T_{d\ge k,k}=k!.
\end{equation}

This fully characterises the initial decay regime (i.e.\ $t<L$), giving us
\begin{equation} \label{eq:decayrates}
  r_1^{(k)} 
  = \begin{cases}
    2  \ln 2,& \text{\ref{CaseA1}},\\
    2k \ln 2 - \ln C_k,& \text{\ref{CaseA2}, $d=2$},\\
    2k \ln d- \ln T_{d,k},& \text{\ref{CaseA2}, arbitrary $d$}.
  \end{cases}
\end{equation}
This is consistent with Fig.~\ref{fig:highermoms}, where the initial decay for Case~\ref{CaseA1} is independent of $k$, and it reproduces the $d=k=2$ decay rate in Fig.~\ref{fig:explicitk2} for~\ref{CaseA2}. For qubits we immediately observe that for an arbitrary $k\ge 1$ we have
\begin{equation}
  \left.r^{(k)}_1\right|_{\text{(a.2)},d=2}\ge
  \left.r^{(1)}_1\right|_{\text{(a.2)},d=2}=
  \left.r^{(k)}_1\right|_{\text{(a.1)}}.
\end{equation}

Note that for $k=1$, we can simplify the diagram in Fig.~\ref{fig:FkDiagram} for any time as $\mathcal{D}_1$ becomes a projector
\begin{equation}\label{eq:defD1Proj}
  \mathcal{D}_1=\ketbra*{\circleSA_1}{\circleSA_1},
\end{equation}
where $\ket{\circleSA_1}$ denotes a (normalised) vectorised identity operator
\begin{equation}
  \ket*{\circleSA_1}=\frac{1}{\sqrt{d}}\sum_{j=1}^{d}\ket{j j}.
\end{equation}
Graphically, we represent Eq.~\eqref{eq:defD1Proj} as 
\begin{equation}
  \begin{tikzpicture}[baseline={([yshift=-0.6ex]current bounding box.center)},scale=0.5]
    \RboundStr{0}{0}{colSt}{1}
  \end{tikzpicture}=
  \begin{tikzpicture}[baseline={([yshift=-0.6ex]current bounding box.center)},scale=0.5]
    \nctgridLine{0}{-0.75}{0}{-0.25}
    \nctgridLine{0}{0.75}{0}{0.25}
    \circle{0}{0.25}
    \circle{0}{-0.25}
  \end{tikzpicture}.
\end{equation}
Furthermore, the unitarity of $U$ implies 
\begin{equation}\label{eq:conditionUnitaryk1}
  \begin{tikzpicture}[baseline={([yshift=-0.6ex]current bounding box.center)},scale=0.5]
    \prop{0}{0}{FcolU}{1}
    \circle{-0.5}{-0.5}
    \circle{0.5}{-0.5}
  \end{tikzpicture}=
  \begin{tikzpicture}[baseline={([yshift=-0.6ex]current bounding box.center)},scale=0.5]
    \draw[white] (0.75,0.75) rectangle (-0.75,-0.75);
    \nctgridLine{-0.5}{0.75}{-0.25}{0.25}
    \nctgridLine{0.5}{0.75}{0.25}{0.25}
    \circle{-0.25}{0.25}
    \circle{0.25}{0.25}
  \end{tikzpicture},\qquad
  \begin{tikzpicture}[baseline={([yshift=-0.6ex]current bounding box.center)},scale=0.5]
    \prop{0}{0}{FcolU}{1}
    \circle{-0.5}{0.5}
    \circle{0.5}{0.5}
  \end{tikzpicture}=
  \begin{tikzpicture}[baseline={([yshift=-0.6ex]current bounding box.center)},scale=0.5]
    \draw[white] (0.75,0.75) rectangle (-0.75,-0.75);
    \nctgridLine{-0.5}{-0.75}{-0.25}{-0.25}
    \nctgridLine{0.5}{-0.75}{0.25}{-0.25}
    \circle{-0.25}{-0.25}
    \circle{0.25}{-0.25}
  \end{tikzpicture},
\end{equation}
while the dual-unitary condition gives,
\begin{equation}\label{eq:conditionDualUnitaryk1}
  \begin{tikzpicture}[baseline={([yshift=-0.6ex]current bounding box.center)},scale=0.5]
    \prop{0}{0}{FcolU}{1}
    \circle{0.5}{0.5}
    \circle{0.5}{-0.5}
  \end{tikzpicture}=
  \begin{tikzpicture}[baseline={([yshift=-0.6ex]current bounding box.center)},scale=0.5]
    \nctgridLine{-0.75}{-0.75}{-0.25}{-0.25}
    \nctgridLine{-0.75}{0.75}{-0.25}{0.25}
    \circle{-0.25}{-0.25}
    \circle{-0.25}{0.25}
  \end{tikzpicture},\qquad
  \begin{tikzpicture}[baseline={([yshift=-0.6ex]current bounding box.center)},scale=0.5]
    \prop{0}{0}{FcolU}{1}
    \circle{-0.5}{0.5}
    \circle{-0.5}{-0.5}
  \end{tikzpicture}=
  \begin{tikzpicture}[baseline={([yshift=-0.6ex]current bounding box.center)},scale=0.5]
    \nctgridLine{0.75}{-0.75}{0.25}{-0.25}
    \nctgridLine{0.75}{0.75}{0.25}{0.25}
    \circle{0.25}{-0.25}
    \circle{0.25}{0.25}
  \end{tikzpicture}.
\end{equation}
Considering now the diagram on the l.h.s.\ of Fig.~\ref{fig:FkDiagram}
for $t>L$, we can first replace the dissipators $\mathcal{D}_1$ with the projector
to $\ket{\circleSA_1}$, and then repeatedly apply
Eq.~\eqref{eq:conditionDualUnitaryk1}, and~\eqref{eq:conditionUnitaryk1} to completely factorise the
diagram obtaining $d^{-2L}$. Combining it with
the result for $t\le L$ we therefore have
\begin{equation}
  F_{t}^{(1)}=\begin{cases}
    d^{-2t},\quad& t\le L,\\
    d^{-2L},& t>L.
  \end{cases}
\end{equation}
Instead, for general ${k>1}$ dual-unitarity does not lead to direct simplifications when time is large compared to the system size. This is expected, as in this regime the dual-unitary dynamics are BQP-complete~\cite{suzuki2022computational}. To gain more analytical insight, we therefore increase the amount of randomness, and consider Case~\ref{CaseB}.

\subsubsection{Case~\ref{CaseB}}
\label{sec:DUcaseB}

Let us now consider Case~\ref{CaseB}, i.e., random onsite matrices everywhere, for dual unitary gates. We begin by focusing on $k=2$ and then move to $k>2$. The time-evolution operator $\mathcal{B}_2$ now consists of averaged gates $W$, which can be understood as the blue gates in Eq.~\eqref{eq:defBlueGate}, decorated with one-site projectors $\mathcal{D}_2$
\begin{equation}\label{eq:averagedDUorangeGate}
  W=\begin{tikzpicture}[baseline={([yshift=-0.6ex]current bounding box.center)},scale=0.5]
    \prop{0}{0}{colSt}{2}
  \end{tikzpicture}
  =
  \begin{tikzpicture}[baseline={([yshift=-0.6ex]current bounding box.center)},scale=0.5]
    \nctgridLine{-1}{-1}{1}{1}
    \nctgridLine{1}{-1}{-1}{1}
    \prop{0}{0}{FcolU}{2}
    \RboundNL{0.65}{0.65}{colSt}{2}
    \RboundNL{0.65}{-0.65}{colSt}{2}
    \RboundNL{-0.65}{-0.65}{colSt}{2}
    \RboundNL{-0.65}{0.65}{colSt}{2}
  \end{tikzpicture}.
\end{equation}
This gives the following diagrammatic expression for $\mathcal{B}_2$
\begin{equation}\label{eq:defB2duCaseB}
  \mathcal{B}_2=
  \begin{tikzpicture}[baseline={([yshift=-0.6ex]current bounding box.center)},scale=0.5]
    \def\X{10}
      \foreach \x in {4,6,...,\X}{\prop{\x-4}{0}{colSt}{2}}
      \foreach \x in {2,4,...,\X}{\prop{\x-3}{1}{colSt}{2}}
    \Lbound{-1.5}{-0.5}
    \nRbound{\X-2.5}{-0.5}
    \draw[decorate,decoration={brace}] (-1.75,2-0.25) -- (\X-2.25,2-0.25) node[midway,yshift=7.5pt] {$L$};
  \end{tikzpicture}.
\end{equation}

The averaged gate $W$ is projected to the subspace spanned by $\ket{\circleSA}$ and $\ket{\squareSA}$ defined as
\begin{equation}
  \begin{aligned} \label{eq:avgDUbasis}
    \ket{\circleSA_2} &= \frac{1}{d} \sum_{s_1,r_1,s_2,r_2} \delta_{s_1,r_1} \delta_{s_2,r_2} \ket{s_1,r_1,s_2,r_2}, \\
    \ket{\squareSA_2} &= \frac{1}{d} \sum_{s_1,r_1,s_2,r_2} \delta_{s_1,r_2} \delta_{s_2,r_1} \ket{s_1,r_1,s_2,r_2},\\
    \ket{\circleSAblack_2} &=  \frac{d \ket{\squareSA_2} - \ket{\circleSA_2}}{\sqrt{d^2-1}},
  \end{aligned}
\end{equation}
and takes the following form in the (orthonormal) basis of $\{\ket{\circleSA_2},\ket{\circleSAblack_2}\}$~\cite{foligno2022growth}
\begin{equation} 
\label{eq:avgDU}
  W = \begin{bmatrix}\vspace{0.1em}
    1 & 0 & 0 & 0 \\
    0 & 0 & 1-p & \frac{p}{\sqrt{d^2-1}} \\
    0 & 1-p & 0 & \frac{p}{\sqrt{d^2-1}} \\
    0 & \frac{p}{\sqrt{d^2-1}} & \frac{p}{\sqrt{d^2-1}} & 1-\frac{2p}{d^2-1}
  \end{bmatrix}.
\end{equation}
Here the free parameter $p$ is the entangling power~\cite{Zanardi_2000,foligno2022growth,aravinda2021from,rather2020creating,RatherConstruction2022}, defined as the average linear entropy generated by the gate when acting on Haar-random product states. In particular, as anticipated in Sec.~\ref{sec:numericalsurvey}, for a DU gates acting on qubits it can be expressed as~\cite{bertini2020scrambling} 
\begin{equation}
  p = \frac{2}{3} \cos (2J)^2,
\end{equation}
where $J$ is the parameter of the dual-unitary gate appearing in Eq.~\eqref{eq:staticgates}.

\paragraph{Late-time regime.}
\label{sec:latetimecaseb}

Let us first consider times $t$ that are large compared to the system size $L$. In this regime the dynamics of $\ket{\smash{\rho_{t}^{(2)}}}$ is governed by the leading subleading eigenvalue of $\mathcal{B}_2$. First let us note that both $\ket*{\circleSA_2\,\circleSA_2}$ and $\ket*{\squareSA_2\,\squareSA_2}$ are invariant under the action of $W$, which implies that $\mathcal{B}_2$ has two eigenvectors corresponding to eigenvalue $1$: $\ket*{\circleSA_2}^{\otimes 2L}$ and $\ket*{\squareSA_2}^{\otimes 2L}$. Property~\ref{prop:p1} guarantees that the latter are the only ones. 

\begin{figure}
  \begin{tikzpicture}[baseline={([yshift=-0.6ex]current bounding box.center)},scale=0.5]
    \def\Y{10}
    \def\X{6}
    \fill[rounded corners=1,fill=colSt] (0,0) rectangle (\X,\Y);
    \pgfmathsetmacro{\old}{7}
    \def\coorsL{{\old,0},{7,1},{8,2},{9,3},{8,4},{9,5},{8,6},{7,7},{6,8},{7,9},{6,10},{5,11},{4,12},{3,13},{4,14},{5,15},{6,16},{5,17},{6,18},{5,19},{4,20}}
    \foreach \coors in \coorsL{
      \pgfmathsetmacro{\L}{{\coors}[0]}
      \pgfmathsetmacro{\dy}{{\coors}[1]}
      \foreach \x in {0,...,\L}{\circle{0.25+0.5*\x}{\Y-0.5*\dy+0.25}}
      \foreach \x in {\L,...,10}{\square{0.75+0.5*\x}{\Y-0.5*\dy+0.25}}
    }
    \newcommand{\nold}{}
    \pgfmathsetmacro{\nold}{\old}
    \draw[thick,colLines,rounded corners=1,fill=colSt,fill opacity=0.5] (0,0) rectangle (\X,\Y);
    \foreach \coors in \coorsL{
      \pgfmathsetmacro{\L}{{\coors}[0]}
      \pgfmathsetmacro{\dy}{{\coors}[1]}
      \draw[thick,red] ({0.5+0.5*\nold},{\Y-0.5*\dy+0.5}) -- ({0.5+0.5*\L},{\Y-0.5*\dy+0.5});
      \draw[thick,red] ({0.5+0.5*\L},{\Y-0.5*\dy+0.5}) -- ({0.5+0.5*\L},{\Y-0.5*\dy});
      \pgfmathsetmacro{\old}{{\coors}[0]}
      \global\let\nold=\old
    }

  \end{tikzpicture}
  \caption{\label{fig:MembranePicture}Schematic representation of the entanglement-membrane picture. The frame potential is mapped onto a partition function of an effective Ising model, whose leading-order contribution comes from the length of the domain wall (red) between the two states (here represented by circles and squares).}
\end{figure}

To find the subleading eigenvalue $\lambda_1$ we follow the entanglement membrane approach of Refs.~\cite{jonay2018coarse, zhou2020entanglement} (see also Ref.~\cite{hunterjones2019unitary} for an application of this approach to frame potentials). The idea is that one can equivalently think of the frame potential as the partition function of an Ising-like spin model whose local degrees of freedom are $\circleSA_2$ and $\squareSA_2$ (because of the form in Eq.~\eqref{eq:averagedDUorangeGate} of the averaged local gate this correspondence is exact), see Fig.~\ref{fig:MembranePicture}. Next one observes that, because the spin model is always in the ordered phase, the leading contribution to the partition function comes from configurations involving a single domain of $\circleSA_2$ or $\squareSA_2$. Going back to the algebraic expression 
\begin{equation}
  F^{(2)}_t = \mel*{\rho_0^{(2)}}{\mathcal{B}_2^{2t}}{\rho_0^{(2)}}, 
\end{equation}
we see that the leading configurations are obtained by plugging a projector over $\ket*{\circleSA_2}^{\otimes 2L}$ (or on $\ket*{\squareSA_2}^{\otimes 2L}$) after each application of the matrix $\mathcal{B}_2$, i.e.,  writing the above equation with the replacement
\be
\mathcal{B}_2 \mapsto  \ketbra*{\circleSA_2}^{\otimes 2L}\mathcal{B}_2 \ketbra*{\circleSA_2}^{\otimes 2L}
\ee
or 
\be
\mathcal{B}_2 \mapsto  \ketbra*{\squareSA_2}^{\otimes 2L}\mathcal{B}_2 \ketbra*{\squareSA_2}^{\otimes 2L}.
\ee
This does indeed lead to the dominant contributions identified by Property~\ref{prop:p1}. 

The question is then what are the dominant subleading configurations. In Haar random circuits is turns out that the latter are simply given by those involving two different domains: one of $\circleSA_2$ and one of $\squareSA_2$, which in the algebraic formulation can be obtained by the replacement $\mathcal{B}_2 \mapsto P_{1} \mathcal{B}_2 P_{1}$, where $P_{1}$ is the projector on the space spanned by 
\begin{equation}
\mathcal{V}_1 = \mathrm{Span}\{\ket*{\circleSA_2}^{\otimes_j-1}\otimes \ket*{\circleSAblack_2}\otimes\ket*{\squareSA_2}^{\otimes_{2L-j}}\}_{j=1}^{2L-1}, 
\end{equation}
where we have written the states in such a way that they are orthogonal.

\begin{figure}[h!]
  \centering
  \includegraphics[width=0.96\linewidth]{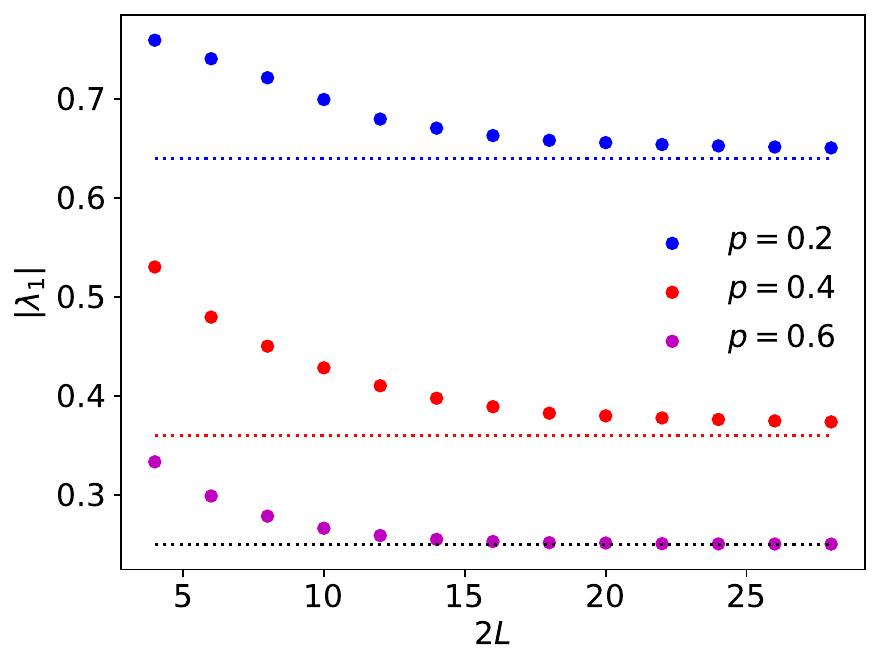}
  \caption{$|\lambda_1|$ vs $L$ for $k=2$ in Case~\ref{CaseB} DU circuits. Horizontal lines correspond to Eq.~\eqref{eq:lambda1DU} and is for $d=2$.}
  \label{fig:gaps}
\end{figure}

In the case of structured circuits, however, the one described above are not the only relevant subleading configurations: Ref.~\cite{Jonay2024} has shown that also `magnon excitations' where one has a single $\circleSA_2$ in a domain of $\squareSA_2$ (and vice versa) can give a non-negligible contribution. In the algebraic setting this means that one should also consider $\mathcal{B}_2 \mapsto P_{1} \mathcal{B}_2 P_{1}$, where $P_{2}$ is the projector on the space spanned by  
\be
\mathcal{V}_2 = \mathrm{Span} (\{\ket*{\circleSA_2}^{\otimes_j}\otimes\ket*{\circleSAblack_2}\otimes  \ket*{\circleSA_2}^{\otimes_{2L-j-1}}\}_{j=0}^{2L-1}).
\ee 
Even though the projectors  $P_{1}$ and $P_2$ are not orthogonal their product becomes increasingly more suppressed in the limit of large sizes the magnitude of dominant subleading eigenvalue of $\mathcal{B}_2$, $|\lambda_1|$, is well approximated by either the spectral radius of $P_1 \mathcal{B}_2 P_1$ or $P_2 \mathcal{B}_2 P_2$. The first one is found to be $d^{-2}$~\cite{zhou2020entanglement}, while the second is the square of the leading eigenvalue of the channel
\be
\label{eq:channelRho2Def}
  \begin{tikzpicture}[baseline={([yshift=-0.6ex]current bounding box.center)},scale=0.5]
    \prop{0}{0}{colSt}{2}
    \circle{-0.5}{0.5}
    \circle{0.5}{-0.5}
 \end{tikzpicture},
\ee
which turns out to be $(1-p)^2$. This gives
\begin{equation}\label{eq:lambda1DU}
  |\lambda_1| \approx \mathrm{max}\{d^{-2}, (1-p)^2\}.
\end{equation}
This prediction is expected to hold in the limit of large $L$, but as we see from numerical test in Fig.~\ref{fig:gaps}, it works well already for moderate system sizes. 

Putting all together we finally find 
\be
\label{eq:Deltascalinglatetimek2}
 \Delta_{2}^{(2)}\sim d^{4 L} \mathrm{max}\{d^{-4t}, (1-p)^{4t}\}. 
\ee
Recalling Eq.~\eqref{eq:designtime}, we therefore have that for ${p\geq  1- 1/d}$ an $\epsilon$-approximate $2$-design is attained in a time 
\be
\tau^{(\epsilon)}_2 \simeq L - \frac{\log \epsilon}{4\ln(d)}\,,
\ee
which we argue is the shortest possible in brickwork circuit.

\paragraph{Early-time regime.} Let us now examine the case in which time is short compared to the system size, more precisely, $L\ge 2t$. In this case we expect to see no contribution from the subspace $\mathcal{V}_2$. To see this, consider the overlap between a vector from $\mathcal{V}_2$ and $\ket{\smash{\rho_{t}^{(2)}}}$
\begin{equation}
  O_{x,t}=\braket*{\underbrace{\circleSA_2\ldots\circleSA_2}_{\phantom{L}x\phantom{L}}\,\circleSAblack_2\,\underbrace{\circleSA_2\ldots\circleSA_2}_{2L-x-1}}{\rho_t^{(2)}}.
\end{equation}
Setting the initial state $\ket*{\rho_0^{(2)}}$ to be a product of pairs as in Eq.~\eqref{eq:diagramBellPairs}, we can express the overlap diagrammatically as
\begin{equation}
  \mkern-6mu
  O_{x,t}=\mkern-8mu
  \begin{tikzpicture}[baseline={([yshift=-1.9ex]current bounding box.center)},scale=0.5]
    \def\X{12}
    \def\Y{6}
    \node at (-2.125,0.5*\Y) {$\displaystyle \frac{1}{d^{4 L}}$};
    \foreach \t in {2,4,...,\Y}{
      \foreach \x in {4,6,...,\X}{\prop{\x-4}{\t-2}{colSt}{2}}
      \foreach \x in {2,4,...,\X}{\prop{\x-3}{\t-1}{colSt}{2}}
    }
    \foreach \x in {2,4,...,\X}{\solstate{\x-3.5}{-0.5}}
    \foreach \t in {2,4,...,\Y}{\Lbound{-1.5}{-2.5+\t}}
    \foreach \t in {2,4,...,\Y}{\nRbound{\X-2.5}{-2.5+\t}}
    \foreach \x in {2,4,...,\X}{\circle{\x-2.5}{\Y-0.5}};
    \foreach \x in {2,4}{\circle{\x-3.5}{\Y-0.5}};
    \foreach \x in {8,10,...,\X}{\circle{\x-3.5}{\Y-0.5}};
    \foreach \x in {6}{\circleBlack{\x-3.5}{\Y-0.5}};
    \draw[decorate,decoration={brace}] (\X-2.125,\Y-0.75) -- (\X-2.125,-0.5) node[midway,right] {$t$};
    \draw[decorate,decoration={brace}] (-1.75,\Y) -- (1.75,\Y) node[midway,above] {$\lfloor \frac{x}{2}\rfloor$};
    \draw[decorate,decoration={brace}] (4.25,\Y) -- (\X-2.25,\Y) node[midway,above] {$\lfloor\frac{2L-x-1}{2}\rfloor$};
  \end{tikzpicture}.
  \mkern-6mu
\end{equation}
The diagram can be simplifying by noting that $\ket*{\circleSA\,\circleSA}$ is invariant under $W$, which is diagrammatically expressed as
\begin{equation}
\label{eq:unitarityfolded}
  \begin{tikzpicture}[baseline={([yshift=-0.6ex]current bounding box.center)},scale=0.5]
    \prop{0}{0}{colSt}{2}
    \circle{-0.5}{-0.5}
    \circle{0.5}{-0.5}
  \end{tikzpicture}=
  \begin{tikzpicture}[baseline={([yshift=-0.6ex]current bounding box.center)},scale=0.5]
    \nRbound{-0.5}{-0.5}
    \Lbound{0.5}{-0.5}
    \circle{-0.5}{-0.5}
    \circle{0.5}{-0.5}
  \end{tikzpicture},\qquad
  \begin{tikzpicture}[baseline={([yshift=-0.6ex]current bounding box.center)},scale=0.5]
    \prop{0}{0}{colSt}{2}
    \circle{-0.5}{0.5}
    \circle{0.5}{0.5}
  \end{tikzpicture}=
  \begin{tikzpicture}[baseline={([yshift=-0.6ex]current bounding box.center)},scale=0.5]
    \nRbound{-0.5}{-0.5}
    \Lbound{0.5}{-0.5}
    \circle{-0.5}{0.5}
    \circle{0.5}{0.5}
  \end{tikzpicture},
\end{equation}
and observing that the dual unitarity of the original gate gives the following two additional relations satisfied by $W$
\begin{equation}
\label{eq:spaceunitarityfolded}
  \begin{tikzpicture}[baseline={([yshift=-0.6ex]current bounding box.center)},scale=0.5]
    \prop{0}{0}{colSt}{2}
    \circle{-0.5}{-0.5}
    \circle{-0.5}{0.5}
  \end{tikzpicture}=
  \begin{tikzpicture}[baseline={([yshift=-0.6ex]current bounding box.center)},scale=0.5]
    \solstate{-0.5}{0.5}
    \solstateD{-0.5}{-0.5}
    \circle{-0.5}{-0.5}
    \circle{-0.5}{0.5}
  \end{tikzpicture},\qquad
  \begin{tikzpicture}[baseline={([yshift=-0.6ex]current bounding box.center)},scale=0.5]
    \prop{0}{0}{colSt}{2}
    \circle{0.5}{-0.5}
    \circle{0.5}{0.5}
  \end{tikzpicture}=
  \begin{tikzpicture}[baseline={([yshift=-0.6ex]current bounding box.center)},scale=0.5]
    \solstate{-0.5}{0.5}
    \solstateD{-0.5}{-0.5}
    \circle{0.5}{-0.5}
    \circle{0.5}{0.5}
  \end{tikzpicture}.
\end{equation}
This immediately gives
\begin{equation}
  \left.O_{j,t}\right|_{j\ge 2t \text{ or } j<2(L-t)}
=\braket*{\circleSAblack_2}{\circleSA_2}=0.
\end{equation}
Thus whenever $L\ge 2t$ the sector corresponding to the subspace $\mathcal{V}_2$ cannot contribute to the decay of $\Delta_{2}^{(2)}$ and for large $t$ we asymptotically expect
\begin{equation} \label{eq:DUearlydecay}
  \left.\Delta_2^{(2)}\right|_{t\le L/2} \sim d^{-2t}.
\end{equation}
In the case $(1-p)^2>d^{-2}$ then we expect to see the relaxation with two different slopes, first $4 \log d$ and then $-4 \log (1-p)$. Instead for $(1-p)^2 \geq d^{-2}$ we expect $\Delta_{2}^{(2)}$ to decay as $d^{-4t}$ in both time regimes and for all entangling powers above the threshold. This is in agreement with the numerical findings reported in Fig.~\ref{fig:AVGDUdynamics}. 

\paragraph{Higher designs: ${k>2}$.}\label{sec:DUkg2} The main insight of Ref.~\cite{Jonay2024}, i.e, that the leading subleading eigenvalue of $ \mathcal{B}_2$ is given by either the `magnon configurations' or the domain walls  (cf. Sec.~\ref{sec:latetimecaseb}), is not limited to $k=2$ and can be applied to arbitrary designs ${k>2}$. Namely, to find the dominant sub-leading eigenvalue of 
\begin{equation}\label{eq:defBkduCaseB}
  \mathcal{B}_k=
  \begin{tikzpicture}[baseline={([yshift=-0.6ex]current bounding box.center)},scale=0.5]
    \def\X{10}
      \foreach \x in {4,6,...,\X}{\prop{\x-4}{0}{colSt}{k}}
      \foreach \x in {2,4,...,\X}{\prop{\x-3}{1}{colSt}{k}}
    \Lbound{-1.5}{-0.5}
    \nRbound{\X-2.5}{-0.5}
    \draw[decorate,decoration={brace}] (-1.75,2-0.25) -- (\X-2.25,2-0.25) node[midway,yshift=7.5pt] {$L$};
  \end{tikzpicture},
\end{equation}
where we introduced  
\begin{equation}\label{eq:averagedDUorangeGatek}
\begin{tikzpicture}[baseline={([yshift=-0.6ex]current bounding box.center)},scale=0.5]
    \prop{0}{0}{colSt}{k}
  \end{tikzpicture}
    =
  \begin{tikzpicture}[baseline={([yshift=-0.6ex]current bounding box.center)},scale=0.5]
    \nctgridLine{-1}{-1}{1}{1}
    \nctgridLine{1}{-1}{-1}{1}
    \prop{0}{0}{FcolU}{k}
    \RboundNL{0.65}{0.65}{colSt}{k}
    \RboundNL{0.65}{-0.65}{colSt}{k}
    \RboundNL{-0.65}{-0.65}{colSt}{k}
    \RboundNL{-0.65}{0.65}{colSt}{k}
  \end{tikzpicture}.
\ee 
The main additional difficulties occurring in the general case are two: (i) the domain wall subspace $\mathcal{V}_1$ is more complicated and involves domain walls among any possible `permutation states' $\ket{\sigma_{j}}$, defined by the following matrix elements in the computational basis of $2k$ qudits ($\ket{s_1 r_1\ldots s_k r_k}_{s_j,r_j=1}^d$)
\be
\braket{s_1 r_1\ldots s_{k} r_k}{\sigma_{j}} = \prod_{m=1}^{k}
    \delta_{s_{m}, r_{\sigma(m)}}\,.
\label{eq:coefficientsperm}
\ee 
(ii) the leading eigenvalue, $\rho_k$, in the magnon subspace $\mathcal{V}_2$, which was determined in Ref.~\cite{Jonay2024} as the dominant sub-leading eigenvalue of 
  \begin{equation}\label{eq:channelRhoKDef}
  \begin{tikzpicture}[baseline={([yshift=-0.6ex]current bounding box.center)},scale=0.5]
    \prop{0}{0}{colSt}{k}
    \circle{-0.5}{0.5}
    \circle{0.5}{-0.5}
 \end{tikzpicture},
  \end{equation}
squared, is not just given by $(1-p)^2$. 

In a random DU circuit, however, it is still true that the leading eigenvalue of $P_1 \mathcal{B}_2 P_1$ continues to be asymptotically given by $d^{-2}$~\cite{zhou2020entanglement}. Therefore Eq.~\eqref{eq:lambda1DU} should be replaced by 
\be
 |\lambda_1| \approx \mathrm{max}\{d^{-2}, \rho_k\}.
\label{eq:eigenvaluegenk}
\ee
This equation suggests that the family of DU circuits with the fastest speed of $k$-design preparation should depend on $k$. In fact, since one can easily show that $\rho_k\geq (1-p)^2$, this family should shrink by increasing $k$. Finding the explicit $k$ dependence is difficult as there is no general way to determine $\rho_k$ analytically for arbitrary values of $k$. The latter, however, can be found for certain specific classes of DU gates. For example, considering \emph{perfect tensors}~\cite{pastawski2015holographic, helwig2012absolute, helwig2013absolutelymaximallyentangledqudit, facchi2008maximally, goyeneche2015absolutely} that, besides the dual-unitarity conditions 
\begin{equation}
\label{eq:unitarityfoldedk}
  \begin{tikzpicture}[baseline={([yshift=-0.6ex]current bounding box.center)},scale=0.5]
    \prop{0}{0}{colSt}{k}
    \circle{-0.5}{-0.5}
    \circle{0.5}{-0.5}
  \end{tikzpicture}=
  \begin{tikzpicture}[baseline={([yshift=-0.6ex]current bounding box.center)},scale=0.5]
    \nRbound{-0.5}{-0.5}
    \Lbound{0.5}{-0.5}
    \circle{-0.5}{-0.5}
    \circle{0.5}{-0.5}
  \end{tikzpicture},\qquad
  \begin{tikzpicture}[baseline={([yshift=-0.6ex]current bounding box.center)},scale=0.5]
    \prop{0}{0}{colSt}{k}
    \circle{-0.5}{0.5}
    \circle{0.5}{0.5}
  \end{tikzpicture}=
  \begin{tikzpicture}[baseline={([yshift=-0.6ex]current bounding box.center)},scale=0.5]
    \nRbound{-0.5}{-0.5}
    \Lbound{0.5}{-0.5}
    \circle{-0.5}{0.5}
    \circle{0.5}{0.5}
  \end{tikzpicture},
\end{equation}
and 
\begin{equation}
\label{eq:spaceunitarityfoldedk}
  \begin{tikzpicture}[baseline={([yshift=-0.6ex]current bounding box.center)},scale=0.5]
    \prop{0}{0}{colSt}{k}
    \circle{-0.5}{-0.5}
    \circle{-0.5}{0.5}
  \end{tikzpicture}=
  \begin{tikzpicture}[baseline={([yshift=-0.6ex]current bounding box.center)},scale=0.5]
    \solstate{-0.5}{0.5}
    \solstateD{-0.5}{-0.5}
    \circle{-0.5}{-0.5}
    \circle{-0.5}{0.5}
  \end{tikzpicture},\qquad
  \begin{tikzpicture}[baseline={([yshift=-0.6ex]current bounding box.center)},scale=0.5]
    \prop{0}{0}{colSt}{k}
    \circle{0.5}{-0.5}
    \circle{0.5}{0.5}
  \end{tikzpicture}=
  \begin{tikzpicture}[baseline={([yshift=-0.6ex]current bounding box.center)},scale=0.5]
    \solstate{-0.5}{0.5}
    \solstateD{-0.5}{-0.5}
    \circle{0.5}{-0.5}
    \circle{0.5}{0.5}
  \end{tikzpicture},
\end{equation}
also fulfil 
\begin{equation}
\label{eq:PTk}
  \begin{tikzpicture}[baseline={([yshift=-0.6ex]current bounding box.center)},scale=0.5]
    \prop{0}{0}{colSt}{k}
    \circle{-0.5}{-0.5}
    \circle{0.5}{0.5}
  \end{tikzpicture}=
  \begin{tikzpicture}[baseline={([yshift=-0.6ex]current bounding box.center)},scale=0.5]
    \draw [thick,colLines] (-0.25,0.25) --  (-0.75,0.75) ;
    \draw [thick,colLines] (0.25,-0.25) --  (0.75,-0.75) ;
    \circle{-0.25}{0.25}
    \circle{0.25}{-0.25}
  \end{tikzpicture},\qquad
  \begin{tikzpicture}[baseline={([yshift=-0.6ex]current bounding box.center)},scale=0.5]
    \prop{0}{0}{colSt}{k}
    \circle{0.5}{-0.5}
    \circle{-0.5}{0.5}
  \end{tikzpicture}=
  \begin{tikzpicture}[baseline={([yshift=-0.6ex]current bounding box.center)},scale=0.5]
    \draw [thick,colLines] (-0.25,-0.25) --  (-0.75,-0.75) ;
    \draw [thick,colLines] (0.25,0.25) --  (0.75,0.75) ;
    \circle{-0.25}{-0.25}
    \circle{0.25}{0.25}
  \end{tikzpicture},
\end{equation}
one immediately has $\rho_k=0$ (and also $p=1$). Tensors fulfilling these properties do not exist for $d=2$ but are known to exist for any $d\geq 3$~\cite{huber2018bounds, rather2022thirty}. As a result, for perfect tensors one expects the same (maximally fast) decay in time for all $\Delta_{2}^{(k)}$, i.e., 
\be
\label{eq:Deltascalinglatetimekg2}
 \Delta_{2}^{(k)}\sim \frac{C}{k!} d^{2 k L} d^{-4t}, \qquad \forall k. 
\ee
For large $L$ and fixed $k$ this gives  
\begin{equation}
\tau^{(\epsilon)}_k \simeq \frac{k}{2}L - \frac{\log \epsilon}{4\ln(d)}\,, 
\end{equation}  
which we argue is the shortest time to achieve an $\epsilon$-approximate $k$ design with a brickwork quantum circuit.

As for the $k=2$ case, Eq.~\eqref{eq:eigenvaluegenk} seems in good agreement with our numerical calculations also for $k=3$, see Tab.~\ref{tab:AVGDUr2}. This is remarkable given the fact that the accessible values of $L$ are substantially smaller in this case than for $k=2$.

\begin{table}
  \centering
  \begin{tabular}{r|l|l|l|l}
    $\hfill L\hfill$ & \hfill 2\hfill\hfill & 
    \hfill 3\hfill\hfill & \hfill 4\hfill\hfill &\hfill 5\hfill\hfill \\\hline
    Perfect tensor & $0.125$ & $0.114574$ & $0.111721$ & $0.111124$ \\
    Hadamard & $0.170067$ & $0.138011$ & $0.123132$ & $0.116489$ \\
    Low entangling & $0.369653$ & $0.322775$ & $0.286805$ & $0.26456$ \\
  \end{tabular}
  \caption{Dominant subleading eigenvalue $|\lambda_1|$ for selected average DU models and $k=3$ and $d = 3$. The gate definitions can be found in the caption of Fig.~\ref{fig:k3dynamics}. For these three cases we have $\rho_3=0$, $\rho_3=0.125$, and~$\rho_3=0.224776$ respectively. In these three cases we also have $\rho_3=(1-p)^2$.
  The value $\lambda_1$ was obtained from a hybrid subspace iteration method to isolate the leading eigen-pairs with $\lambda_{0,q}=1$, followed by an Arnoldi method to isolate $\lambda_{1,q}$. In the low entangling case for $L=5$ the convergence criteria were not met and we report the final eigenvalue magnitude estimate.}
  \label{tab:AVGDUr2}
\end{table}

\section{Conclusions}
\label{sec:conclusions}

In this work we investigated quantum-state-design preparation in local quantum circuits with a minimal degree of randomness. We considered two cases, both in the framework of brickwork circuits of finite size: (a) fixed circuits with a single one-qudit random unitary matrix at the boundary; (b) brickwork circuits with one-qudit random unitary matrices everywhere but fixed interactions. Both cases are characterised by the property that the entangling part of the local gate is \emph{not} random. We obtained two main results. 

First, we found that both the settings considered produce quantum state designs for sufficiently large depths. In fact, if the local gates are sufficiently entangling, they do so \emph{faster} than fully random (Haar random) brickwork circuits. More specifically, we found that if the local gates are dual-unitary there is a direct relationship between the speed of $2$-design preparation and the entangling power~\cite{Zanardi_2000,aravinda2021from,rather2020creating,RatherConstruction2022} of the local gates. The speed of $2$-design preparation grows monotonically with the entangling power for small values of the latter, and, above a critical value, it saturates to its maximal value --- this is a crossover for finite width but approaches a phase transition in the limit of infinite width. As a result, even for moderate values of the entangling power dual-unitary circuits produce $2$-designs faster than Haar random circuits. This finding is in line with the recent observation of Ref.~\cite{claeys2025fock}, crucially, however, our statement pertains to a finite system, rather than a thermodynamically large one. Instead, away from the dual-unitary point the speed of $2$-design preparation depends on both entangling power and dual-unitarity breaking. We argued that the production speed of more general $k$-designs should not only depend on the entangling power, however, we showed that dual-unitary circuits with maximal entangling power --- perfect tensors~\cite{pastawski2015holographic, helwig2012absolute, helwig2013absolutelymaximallyentangledqudit, facchi2008maximally, goyeneche2015absolutely} --- give the optimal $k$-design preparation for any $k$. We also demonstrated  numerically that, for $k=3$, this is the case also for dual-unitary gates in the Hadamard family~\cite{aravinda2021from}. 

Second, we also found that, in contrast to the random circuit setting, the quantum state $k$-design preparation in minimally random quantum circuits occurs in a two-step fashion --- similarly to what happens for purity~\cite{bensa2021fastest, znidaric2022solvable, znidaric2023phantom} and out-of-time-ordered correlators~\cite{bensa2022two, znidaric2023two} --- and provided an analytic description of this phenomenon in the case of dual-unitary circuits. These findings suggest that that the two settings we studied here are more similar to each other than to Haar random circuits, meaning that the setting (b) represents a qualitatively more accurate modelling of non-random dynamics than the fully random circuit.

Combined together, our results indicate that one can significantly improve the preparation of Haar random states in brickwork circuits by considering specific families of gates with a minimal amount of randomness and identify dual-unitary with high entangling power as one such families. In a practical implementation this reduces both the required sample size and the time to run a single sample. In fact, a natural expectation arising from our findings is that this minimally random setting can also be used for the efficient preparation of \emph{unitary} rather than \emph{quantum state} $k$ designs, i.e., of random unitary matrices rather than random states. Indeed, reasoning along the lines of Refs.~\cite{bertini2018exact, bertini2021random}, and \cite{ho2022exact}, one can directly show that our statement about the state becoming random at late times can be turned into one about the evolution operator for $t$ steps becoming a random matrix for large $t$. Moreover, our observations on the convergence time can be directly applied to characterise the time scale of this latter process. 
More generally, by showing that one can replace global randomness with local randomness combined with (deterministic) chaotic dynamics, our results offer a different `practical' perspective on the concept of quantum chaos.

A natural question for future research is to establish rigorously whether this is the case by proving some of the statements that are argued or conjectured here. For example, that the design production speed achieved by perfect tensors is the theoretical maximum achievable in a brickwork circuit and whether one can achieve this speed in the setting (a). Moreover,
it is compelling to ask whether the setting described here is the optimal one (i.e.\ the one requiring the least amount of resources) or can be further improved.

\begin{acknowledgments}
We thank Xhek Turkeshi for useful comments on the manuscript. We acknowledge financial support from the Royal Society through the University Research Fellowship No.\ 201101 (B.\ B.\ and J.\ R.) and the Leverhulme Trust through the Early Career Fellowship No.\ ECF-2022-324 (K.\ K.). 
\end{acknowledgments}

\appendix

\section{Proof of Property~\ref{prop:p1}}
\label{app:IppolitiHotheorem}

Property~\ref{prop:p1} follows by straightforward adaptation of Ippoliti and Ho's Theorem (cf.\ Sec.~3.C in Ref.~\cite{Ippoliti2023}). Let us provide a brief proof considering separately the three Cases~\ref{CaseA1}, \ref{CaseA2}, and~\ref{CaseB1}. 

\subsection{Proof for Case~\ref{CaseA1}}

In Case~\ref{CaseA1} the mixed state $\rho^{(k)}_t$ evolves according to 
\be
\label{eq:channelcase1}
\rho^{(k)}_{t+1} =  \mathcal B_k[\rho_{t}^{(k)} ] = \mathcal{U}^{\otimes k} \mathcal{D}_k[\rho^{(k)}_t ] (\mathcal{U}^\dag)^{\otimes k}
\ee
where $\mathcal{U}$ is written in terms of matrices $U^{(n)}\in\mathcal A_V$ as in Eq.~\eqref{eq:circuit} and 
\be
\label{eq:channel}
\mathcal{D}_k[\cdot] = 4^{-1}
\smashoperator{\sum_{\alpha \in \{I, \sigma^x, \sigma^y, \sigma^z\}}} (\alpha_{2L})^{\otimes k} \cdot (\alpha_{2L})^{\otimes k}.  
\ee
This is very similar to the setting considered in Ref.~\cite{Ippoliti2023}: the only difference is that $\mathcal A_V$ is an open subset of $U(4)$ and not necessarily of the dual-unitary submanifold. Therefore, we need to show that the Lemma in Appendix E of Ref.~\cite{Ippoliti2023}, the one guaranteeing that one can approximate any unitary by a brickwork circuit of gates taken from $\mathcal A_V$, still holds. Repeating the arguments of Ref.~\cite{Ippoliti2023} we then need to show that 
\be
(UW^\dag)_{n,n+1},
\ee
can generate any unitary operation by varying $n\in \mathbb Z_{2L}$, and $U, W\in \mathcal A_V$. Considering $W$ close to $U$ we see that one can then generate any arbitrary infinitesimal unitary transformation on the qubits at position $n$ and $n+1$. Taking then sufficiently high powers one can generate any unitary transformation on the two qubits. Since $n$ is arbitrary the statement immediately follows. Following Ref.~\cite{Ippoliti2023} we then have that the limit  
\be
\rho^{(k)}_{\infty}=\lim_{t\to\infty} \rho^{(k)}_{t},
\ee
exists and commutes with 
\be
({\mathcal U} \sigma^\alpha \mathcal U^\dag)^{\otimes k},
\ee
where ${\mathcal U}$ is an arbitrary unitary matrix in $U(2^{2L})$. These two matrices can be chosen in such a way to generate a complete gate set for each pair of qubits on the chain. Therefore, we conclude that $\rho^{(k)}_{\infty}$ commutes with any unitary matrix of the form ${\mathcal U}^{\otimes k}$, with ${\mathcal U}$ in $U(2^{2L})$. The Schur--Weyl duality~\cite{fulton1991representation, marvian2014generalization} then implies that $\rho^{(k)}_{\infty}=\rho^{(k)}_{H}$.


\subsection{Proof for Case~\ref{CaseA2}}

In Case~\ref{CaseA2} the channel $\mathcal B_k[\cdot]$ can again be written as in Eq.~\eqref{eq:channelcase1}. In this case, however, $d$ is a generic integer $\geq 2$ and $\mathcal{D}_k[\cdot]$ is modified to   
\be
\mathcal{D}_k[\cdot] = \int_{U(d)}  (\alpha)^{\otimes k}\cdot\, (\alpha_{2L}^\dag)^{\otimes k} \mu_H(\alpha_{2L})\,,
\ee
Therefore the discussion in Ref.~\cite{Ippoliti2023} does not directly apply. Specifically, we need to prove again the Lemmas in Appendix D of Ref.~\cite{Ippoliti2023}, those concerning the existence of the limit state and the fact that it commutes with the Krauss operators of the channel. By noting that, however, $\mathcal{D}_k[\cdot]$ projects the state at site $2L$ on the space spanned by $\{P_\sigma\}_{\sigma\in S_k}$, where $S_k$ is the symmetric group of of $k$ elements and 
\be
[P_\sigma]_{\boldsymbol i,\boldsymbol j} = \prod_{p=1}^k \delta_{i_p,j_{\sigma(p)}}, 
\ee
one finds that the arguments of Ref.~\cite{Ippoliti2023} can be straightforwardly repeated and both Lemmas continue to hold. Also the Lemma in Appendix E holds as the arguments of the previous subsection are not restricted to $d=2$. Therefore, we again have that the limit  
\be
\rho^{(k)}_{\infty}=\lim_{t\to\infty} \rho^{(k)}_{t},
\ee
exists and commutes with 
\be
({\mathcal U} u_{2L} \mathcal U^\dag)^{\otimes k},
\ee
where $u$ is an arbitrary unitary matrix in $U(d)$ and ${\mathcal U}$ is an arbitrary unitary matrix in $U(d^{2L})$. These two matrices can be chosen in such a way to generate a complete gate set for each pair of qudits on the chain. Therefore, we again conclude that $\rho^{(k)}_{\infty}$ commutes with any unitary matrix of the form ${\mathcal U}^{\otimes k}$, with ${\mathcal U}$ in $U(d^{2L})$. The Schur--Weyl duality~\cite{fulton1991representation, marvian2014generalization} then implies that $\rho^{(k)}_{\infty}=\rho^{(k)}_{H}$.

\subsection{Proof for Case~\ref{CaseB1}}

The statement for Case~\ref{CaseB1} immediately follows from Case~\ref{CaseA2}. One can first average only on the unitary at the boundary producing the channel studied in the previous subsection. Namely, the $k$-moment state is written as 
\be
\rho^{(k)}_{t} = \int \rho^{(k)}_{t}(\boldsymbol \alpha) \prod_{j=1}^{2L-1} \prod_{\tau=1}^{t}  \mu_H(\alpha^{(j)}_{2\tau-1}) \mu_H(\alpha^{(j)}_{2\tau}),
\ee
where the integrand is a step of evolution of the channel studied in the previous subsection for a given value of the one-site unitaries $\boldsymbol \alpha = (\alpha^{(1)}_1, ...., \alpha^{(2L-1)}_{2t})$. Therefore we readily have   
\be
\rho^{(k)}_{\infty} = \int \rho^{(k)}_{H} \prod_{j=1}^{4Lt-1} \mu_H(\alpha) = \rho^{(k)}_{H}\,. 
\ee

\bibliography{bibliography}  
\end{document}